\documentclass[fleqn,usenatbib]{mnras}

\usepackage{newtxtext,newtxmath}
\usepackage[T1]{fontenc}
\usepackage{ae,aecompl}

\usepackage{graphicx}	% Including figure files
\usepackage{amsmath}	% Advanced maths commands
\usepackage{amssymb}	% Extra maths symbols
\usepackage{adjustbox}
\usepackage{rotating}
\usepackage{hyperref}
\usepackage[flushleft]{threeparttable}
\usepackage{color,soul}

\makeatletter
\newcommand*{\rom}[1]{\expandafter\@slowromancap\romannumeral #1@}
\makeatother

\title[Drivers of the chemical complexity in G+0.693]{Cloud-cloud collision as drivers of the chemical complexity in Galactic Centre molecular clouds}

\author[S.Zeng et al.]{
S. Zeng,$^{1,2,3}$\thanks{E-mail: shaoshan.zeng@riken.jp}
Q. Zhang,$^{2}$
I. Jim\'enez-Serra,$^{4}$
B.Tercero,$^{5,6}$
X. Lu,$^{7}$
\newauthor{
J. Mart\'in-Pintado,$^{4}$
P. de Vicente,$^{6}$
V. M. Rivilla,$^{8}$
and S. Li$^{2,9,10}$}
\\
$^{1}$School of Physics and Astronomy, Queen Mary University of London, Mile End Road, E1 4NS London, UK\\
$^{2}$Center for Astrophysics $\mid$ Harvard $\&$ Smithsonian, 60 Garden Street, Cambridge, MA 02138, USA\\
$^{3}$Star and Planet Formation Laboratory, RIKEN Cluster for Pioneering Research, 2-1, Hirosawa, Wako, Saitama 351-0198, Japan\\
$^{4}$Centro de Astrobiolog\'ia (CSIC-INTA), Carretera de Ajalvir, Km. 4, Torrej\'on de Ardoz, 28850 Madrid, Spain \\
$^{5}$Observatorio Astron\'omico Nacional (OAN-IGN), Calle Alfonso XII, 3, 28014 Madrid, Spain \\
$^{6}$Observatorio de Yebes (IGN), Cerro de la Palera S/N, 19141, Guadalajara, Spain\\
$^{7}$National Astronomical Observatory of Japan, 2-21-1 Osawa, Mitaka, Tokyo, 181-8588, Japan \\
$^{8}$INAF-Osservatorio Astrofisico di Arcetri, Largo Enrico Fermi 5, 50125, Florence, Italy\\
$^{9}$Shanghai Astronomical Observatory, Chinese Academy of Sciences, 80 Nandan Road, Shanghai 200030, China\\
$^{10}$University of Chinese Academy of Sciences, 19A Yuquanlu, Beijing 100049, China\\
}

\date{Accepted XXX. Received YYY; in original form ZZZ}

\pubyear{2020}

\begin{document}
\label{firstpage}
\pagerange{\pageref{firstpage}--\pageref{lastpage}}
\maketitle

\begin{abstract}
G+0.693-0.03 is a quiescent molecular cloud located within the Sagittarius B2 (Sgr B2) star-forming complex. Recent spectral surveys have shown that it represents one of the most prolific repositories of complex organic species in the Galaxy. The origin of such chemical complexity, along with the small-scale physical structure and properties of G+0.693-0.03, remains a mystery. In this paper, we report the study of multiple molecules with interferometric observations in combination with single-dish data in G+0.693-0.03. Despite the lack of detection of continuum source, we find small-scale (0.2 pc) structures within this cloud. The analysis of the molecular emission of typical shock tracers such as SiO, HNCO, and CH$_3$OH unveiled two molecular components, peaking at velocities of 57 and 75 km\,s$^{-1}$. They are found to be interconnected in both space and velocity. The position-velocity diagrams show features that match with the observational signatures of a cloud-cloud collision. Additionally, we detect three series of class \rom{1} methanol masers known to appear in shocked gas, supporting the cloud-cloud collision scenario. From the maser emission we provide constraints on the gas kinetic temperatures ($\sim$30-150 K) and H$_2$ densities (10$^4$-10$^5$ cm$^{-2}$). These properties are similar to those found for the starburst galaxy NGC253 also using class \rom{1} methanol masers, suggested to be associated with a cloud-cloud collision. We conclude that shocks driven by the possible cloud-cloud collision is likely the most important mechanism responsible for the high level of chemical complexity observed in G+0.693-0.03.

\end{abstract}


%%%%%%%%%%%%%%%%%%%%%%%%%%%%%%%%%%%%%%%%%%

\section{Introduction}
The inner $\sim$500 pc of the Milky Way is known as the Central Molecular Zone (or CMZ). It is a heavily processed region and a hotbed of star formation activity, containing the most extreme massive star formation sites \citep{Morris1996} in our Galaxy. A significant part of the molecular gas in the CMZ, however, is found in quiescent giant molecular clouds (GMCs), which exhibit a high level of turbulence (highly supersonic) and little, or non-existent, star formation activity \citep[e.g.][]{Longmore2013,Kauffmann2017,Lu2019a,Lu2019b}. In contrast to GMCs in the Galactic disc, GMCs in the CMZ are exposed to energetic phenomena such as shock waves (due to the high level of turbulence), intense UV radiation fields (from nearby massive stellar clusters such as the Arches and the Quintuplet clusters), X-rays (coming from the central and massive black hole) and enhanced cosmic-ray ionisation rates (\citealp[e.g.][]{Bally1987,Koyama1989,Goto2013,Yusef-Zadeh2013}; \citealp[see also][and references therein]{mills2018}). The physical conditions of the CMZ molecular clouds therefore differ substantially from the rest of the Galaxy. Indeed, the average H$_2$ gas density \citep[$\sim$10$^4$ cm$^{-3}$,][]{Guesten1983,Bally1987a,Rodriguez-fernandez2000} is several orders of magnitude above the average in the Galactic disc, and the gas temperature typically ranges between $\sim$30 to $\sim$150 K \citep{Huettemeister1993,Guesten1985,Ott2014,Ginsburg2016,Krieger2017}. The gas temperature is decoupled from the much lower temperature of dust grains of $\leq$20 K \citep{Rodriguez-fernandez2004,Nagayama2009}.

One molecular cloud that stands out over the others within the CMZ is designated by its Galactic coordinates as G+0.693-0.03 (hereafter G+0.693). Single-dish spectral surveys performed with the IRAM 30\,m and GBT telescopes have revealed a plethora of molecular species including many complex organic molecules (COMs) in the region of G+0.693 \citep{Requena-torres2008,Rivilla2018, Rivilla2019,Zeng2018,Jimenez-Serra2020}. It is believed that G+0.693 represents one of the largest molecular repositories of COMs in our Galaxy \citep{Requena-torres2006,Requena-torres2008}. G+0.693 can be found within the Sagittarius B2 (Sgr B2 hereafter) star-forming cloud, which is known to be one of the most active sites of star formation in our Galaxy. As a whole, Sgr B2 can be distinguished into three different parts: a low-density envelope, a moderate density region, and the most compact, densest molecular regions \citep[as illustrated in][]{Schmiedeke2016}. 

At the centre of the envelope, three sources named according to their relative location in an equatorial coordinate system, Sgr B2(North), Sgr B2(Main), and Sgr B2(South), are positioned along a north-south ridge. Especially, the former two are well-known sites for active star formation that comprise all the typical signposts of massive star formation such as hot cores, molecular masers of H$_2$O, OH, H$_2$CO, CH$_3$OH, and SiO, and ultracompact H \rom{2} regions \citep[see e.g.][and references therein]{Martin-Pintado1999, deVicente2000, Jones2008}. The morphology and kinematic features identified in Sgr B2 have been interpreted as the result of a cloud-cloud collision \citep{Hasegawa1994, Hasegawa2008, Sato2000, Tsuboi2015}, which possibly triggered this intense star-formation activity. 

Considering Sgr B2N is at an earlier stage of star formation than Sgr B2M \citep{Miao1995, deVicente2000}, one might postulate that the star formation activity occurs sequentially from Sgr B2M to north. If this is the case, early star formation activity is also expected in G+0.693 since it is located also at the ridge just only $\sim$55 arcsec northeast away from Sgr B2N. However, signposts of ongoing star formation such as UC H \rom{2} regions, H$_2$O masers, class \rom{2} CH$_3$OH masers, or H$_2$CO masers or even precursors of massive star formation such as dust continuum sources have not yet been detected towards G+0.693 \citep{Ginsburg2018,Lu2019b}. The nature of G+0.693 remains quiescent. Recent star formation of intermediate (2-8 M$\odot$) and high-mass stars (>8 M$\odot$) can however be masked even in high-sensitivity continuum observations by large amounts of dust in deeply embedded molecular clouds, as found for the intermediate mass hot core found in the Cepheus A HW2 system \citep[see e.g.][]{Martin-Pintado2005,Jimenez-Serra2009}. Therefore, if present, high-angular resolution observations of rotational molecular lines are better probes of young massive star formation in the form of intermediate- and high-mass hot molecular cores \citep{Jimenez-Serra2007, Jimenez-Serra2009}. In fact, the presence of hot core-like sources in G+0.693 would explain the high level of chemical complexity found in this source, with over 40 COMs detected including the prebiotic molecules cyanomethanimine and urea \citep{Requena-torres2008, Zeng2018,Rivilla2019,Jimenez-Serra2020}. Up to date, there are more than 200 molecules that have been detected in the ISM of which about one third of them ($\sim$70) are COMs (molecules containing six or more atoms)\footnote{https://cdms.astro.uni-koeln.de/classic/molecules}. G+0.693 is therefore an ideal candidate to improve our understanding of the origin of COMs in the ISM.

Besides, the chemical richness observed in G+0.693 could also be related to the presence of large-scale, low-velocity shocks that populate the Galactic Centre. In low-velocity shocks (with shock velocity of $\sim$20 kms$^{-1}$), dust grains are efficiently sputtered, fully releasing the materials present in the icy mantles (including COMs), and partially eroding grain cores \citep[see e.g.][]{Jimenez-Serra2008}. This is consistent with widespread SiO emission detected across the Galactic Centre \citep{Martin-pintado1997}. Additional support for the shock scenario comes from the detection of class I methanol masers in the region \citep[see][]{Liechti1996, Jones2011}. Usually class I methanol masers are formed in regions of recent star formation associated with outflow activity. It is believed that the physical conditions and the CH$_3$OH abundance required for this type of masers in the material has experienced shock interaction \citep[e.g.][]{Voronkov2006,Cyganowski2009,Pihlstrom2014}.

The origin of a large-scale, low-velocity shock in the quiescent GMC G+0.693 is unknown. Two scenarios have been proposed: 1) the shocks in G+0.693 could be produced by small-scale expanding wind-blown bubbles driven by evolved massive stars \citep{Martin-Pintado1999}; or 2) these shocks are associated with a large-scale cloud-cloud collision, as mentioned earlier. \citet{Henshaw2016} investigated the large-scale kinematics of Sgr B2N complex and identified a conical profile of Sgr B2 in the position-position-velocity space (see their Figure 17). This feature coincides with the location of G+0.693 where two gas streams could be merging into a cloud-cloud collision, consistent with the second hypothesis. Other interpretations are related to the global orbital structure of the molecular gas in the CMZ, or a signature of global collapse (J. Henshaw, private communication).

With the purpose of understanding the origin of COMs in G+0.693 i.e. the main mechanism(s) responsible for the peculiar chemistry observed, we have used interferometric submillimetre observations of several rotational molecular lines to gain insight into the morphology, the kinematics, and the physical properties of G+0.693 at small scales, which have not been studied in detail before. We also report the detection of a series of class $\,$ \rom{1} methanol masers, usually associated with shocks, towards this source by using single-dish telescopes which provide constraints on the physical properties of the shocked gas experiencing the maser amplification. This paper is distributed as follows: Section \ref{Data} describes the data used throughout this study. The main results are depicted and discussed in Section \ref{Results} and \ref{Discussions}. Throughout this study, we adopt a distance to the Galactic Centre of 8.4 kpc \citep{Reid2014}.

%%%%%%%%%%%%%%%%%%%%%%%%%%%%%%%%%%%%%%%%%%

\section{Observations}
\label{Data}
\subsection{Submillimeter Array (SMA) observations}
We used the archival Submillimeter Array (SMA)\footnote{The SMA is a joint project between the Smithsonian Astrophysical Observatory and the Academia Sinica Institute of Astronomy and Astrophysics and is funded by the Smithsonian Institution and the Academia Sinica.} data in the compact configuration from CMZoom survey \citep[project code: 2013B-S091, PIs: C. Battersby and E. Keto;][]{Battersby2017,Battersby2020} and subcompact configuration (project code: 2013B-S055) to study the small-scale morphology and kinematics of G+0.693. The data set is a six-pointing mosaic around the position of G+0.693 observed at the 230\,GHz band in 2014. The primary beam of SMA is $\sim$55$^{\prime\prime}$ at the observed frequency. A total bandwidth of 8\,GHz is covered in two sidebands: rest frequencies of 216.9-220.8\,GHz covered in the lower sideband whilst 228.9-232.9\,GHz covered in the upper sideband. The continuum data were also obtained within the same observation by averaging line-free spectral channels over the 8\,GHz bandwidth. The typical spectral resolution is 0.812 MHz which is equivalent to a velocity resolution of $\sim$1.1 kms$^{-1}$. 

Data calibration was carried out in \textsc{mir}\footnote{\textsc{mir} is an IDL-based package developed to calibrate SMA data (Available at https://www.cfa.harvard.edu/~cqi/mircook.html.)} and \textsc{miriad} \citep{sault1995}. Spectral lines and continuum were subsequently imaged and analysed using \textsc{casa} \citep{Mcmullin2007}. A 1.3 mm continuum map was generated averaging channels free of line emission. The resulting image has a synthesised beam of 4$^{\prime\prime}$.0 $\times$ 3$^{\prime\prime}$7 (equivalent to 0.16 pc $\times$ 0.14 pc) with a position angle of 30$^{\circ}$.2. Each of the detected molecular lines was imaged separately. The line maps have a typical synthesised beam of $\sim$4$^{\prime\prime}$.4 $\times$ 4$^{\prime\prime}$.0 (equivalent to 0.18 pc $\times$ 0.16 pc) with a position angle of 33$^{\circ}$.7, The rms of continuum map and spectral line maps is $\sim$5 mJy beam$^{-1}$ and $\sim$0.2 Jy beam$^{-1}$ per 1.1 km\,s$^{-1}$ channel respectively. 

\subsection{Atacama Pathfinder EXperiment (APEX) data}
For the spectral lines, we also utilised the single-dish data observed with Atacama Pathfinder EXperiment (APEX) \citep{Ginsburg2016} at 216.9-220.9\,GHz which covers the same frequency range as the lower sideband of our SMA observation. However, no single-dish data are available to cover the same frequency range as the upper sideband (228.9-232.9\,GHz). Hence molecules detected in the upper sideband will not be imaged and analysed.

%\subsection{Combination of interferometric and single-dish data}
Although the SMA observations were performed in both compact and subcompact configurations, the data still suffer from missing flux due to limited uv-converage of short baselines of the interferometer. To account for this, we combined the SMA interferometric data with the APEX single-dish data by employing the \textit{feather} task in \textsc{casa}. All the maps presented in this work are not corrected for primary-beam response in order to have a uniform rms level across the maps.

%%%%%%%%%%%%%%%%%%%%%%%%%%%%%%%%%%%%%%%%%%
%%%%%%%%%%%%%%%%%%%%%%%%%%%%%%%%%%%%%%%%%%
\begin{figure*}
    \centering
    \includegraphics[width=0.8\linewidth]{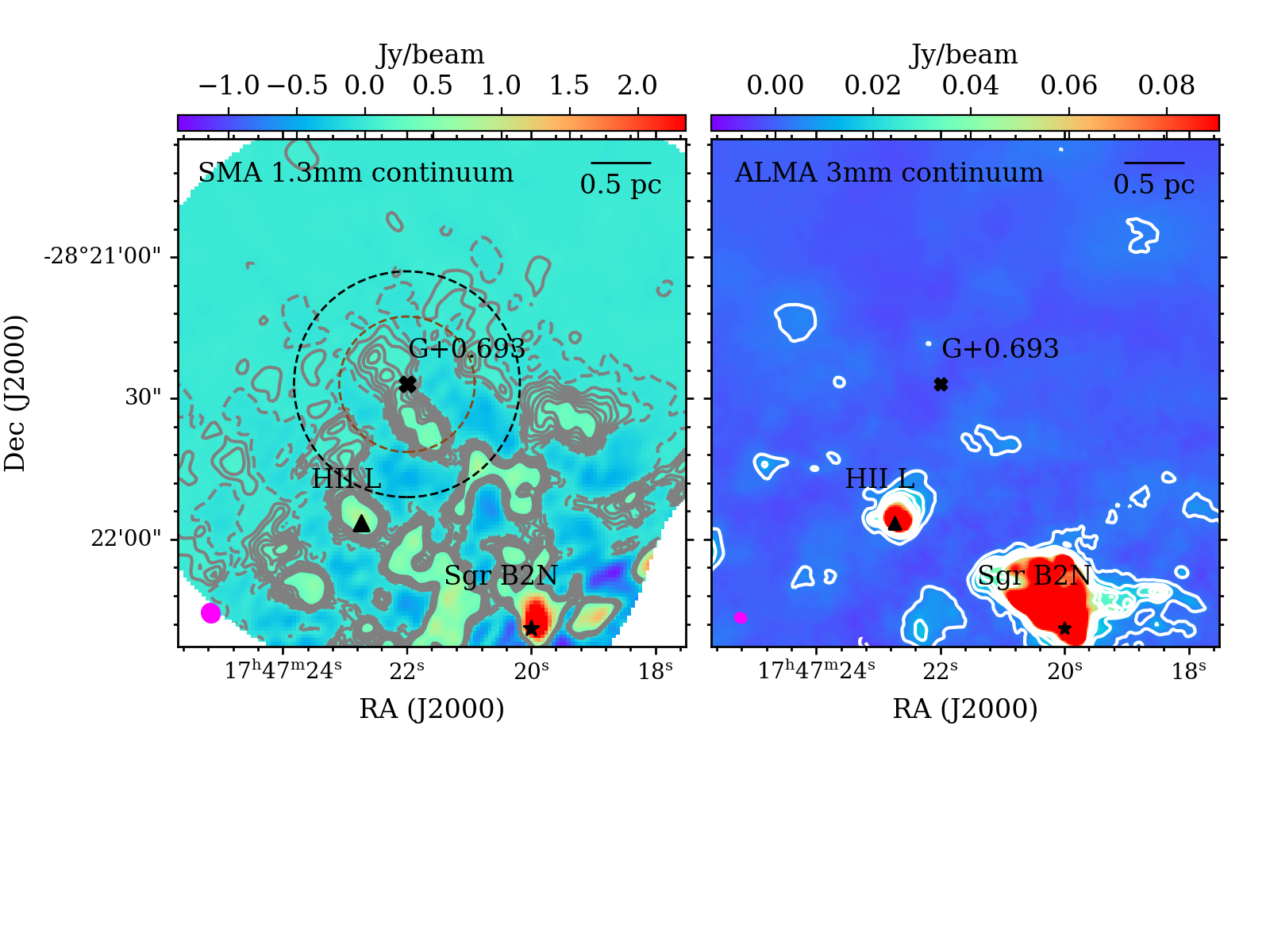}
    \vspace{-2.5cm}
    \caption{Left panel: SMA 1.3 mm continuum map centred at the position of G+0.693. The grey solid contours represent 5$\sigma$-55$\sigma$ levels in steps of 10$\sigma$, where 1$\sigma$=5 mJy/beam whilst grey dashed contour represents -5$\sigma$ level. The SMA synthesised beam, 4$^{\prime\prime}$.0 $\times$ 3$^{\prime\prime}$7 with a position angle of 30$^{\circ}$.2, is shown in the lower-left corner. The black cross, triangle and star symbol indicate the position of G+0.693, H\rom{2} region L, and Sgr B2N respectively. The brown and black dashed circle indicate respectively the largest beam size ($\sim$29$^{\prime\prime}$) of IRAM 30\,m observations and Yebes 40\,m ($\sim$48$^{\prime\prime}$) observations towards G+0.693. Right panel: ALMA 3 mm continuum map from \citet{Ginsburg2018} centred at the position of G+0.693. The white contours represent 5$\sigma$-55$\sigma$ levels in steps of 10$\sigma$, where 1$\sigma$=0.78 mJy/beam. The synthesised beam, 2$^{\prime\prime}$.35 $\times$ 1$^{\prime\prime}$.99 with a position angle of 74$^{\circ}$.6, is shown in the lower left corner. The label for each source is the same as left panel.}
    \label{fig:sma-continuum}
\end{figure*}
%%%%%%%%%%%%%%%%%%%%%%%%%%%%%%%%%%%%%%%%%%
%%%%%%%%%%%%%%%%%%%%%%%%%%%%%%%%%%%%%%%%%%

%%%%%%%%%%%%%%%%%%%%%%%%%%%%%%%%%%%%%%%%%%
%%%%%%%%%%%%%%%%%%%%%%%%%%%%%%%%%%%%%%%%%%

\begin{figure*}
    \centering
    \includegraphics[width=0.8\linewidth]{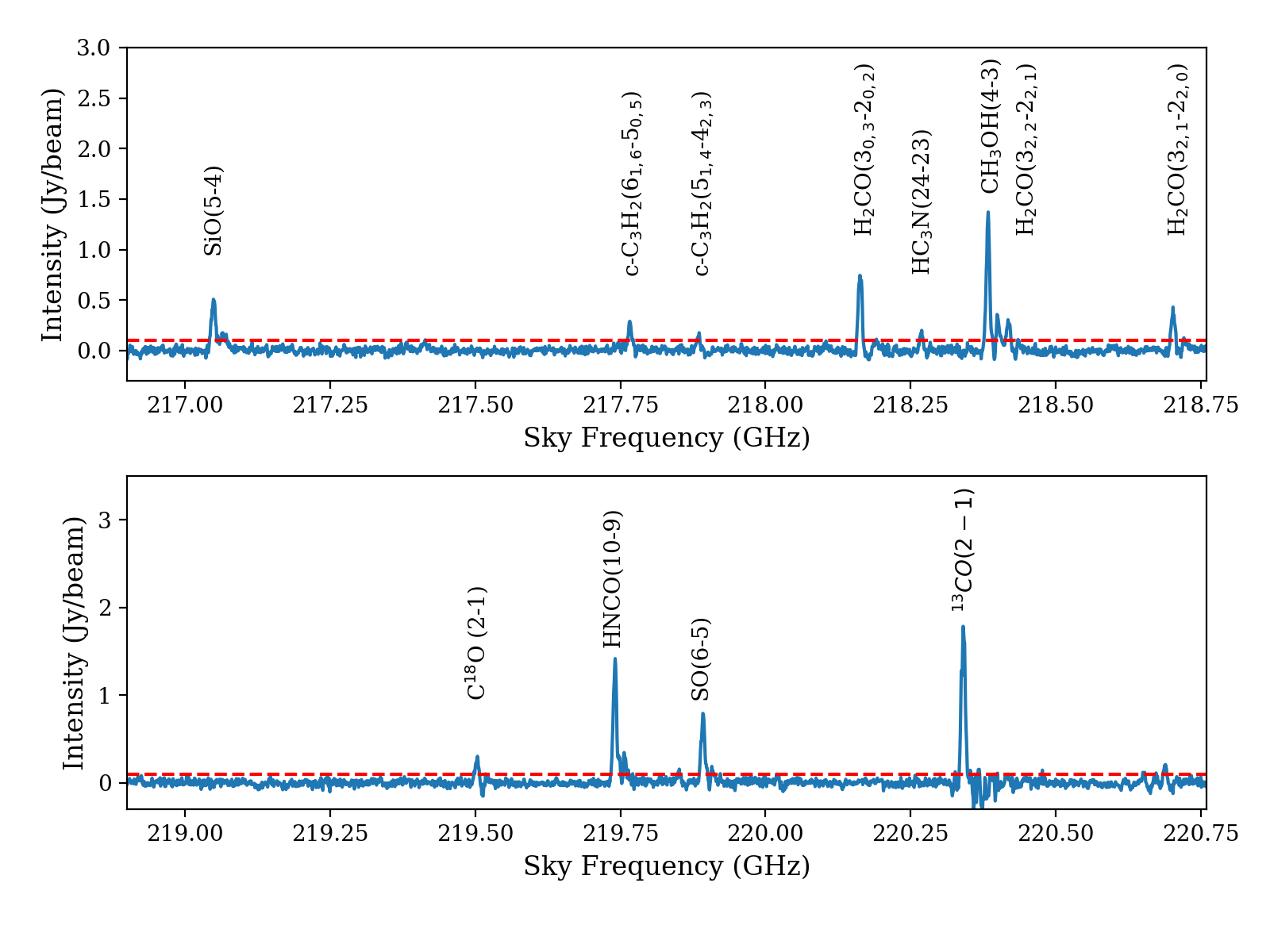}
    \vspace{-0.9cm}
    \caption{Full spectra measured between 216.9-218.8\,GHz (upper panel) and 218.9-220.8\,GHz (lower panel) in the lower sideband of the SMA observations. Detected molecular species are identified and labelled with their name and transition. The horizontal red dashed line marks the 3$\sigma$ level.}
    \label{fig:full-spectra}
\end{figure*}

%%%%%%%%%%%%%%%%%%%%%%%%%%%%%%%%%%%%%%%%%%
%%%%%%%%%%%%%%%%%%%%%%%%%%%%%%%%%%%%%%%%%%

\subsection{Single-dish observations with Yebes 40 m, IRAM 30 m and GBT telescope}

The observations of the class \rom{1} methanol masers at 36.169\,GHz and at 44.069\,GHz were carried out with the Yebes 40\,m telescope\footnote{http://rt40m.oan.es/rt40m$\_$en.php} \citep{deVicente2016} located at 990\,m of altitude in Yebes (Guadalajara, Spain), during 6 observing sessions in February 2020. The equatorial coordinates of the G+0.693 source were \mbox{$\upalpha_{\rm J2000}$\,=\,17$^{\rm h}$\,47$^{\rm m}$\,22$^{\rm s}$}, \mbox{$\updelta_{\rm J2000}$\,=\,$-$28$^{\circ}$\,21$'$\,27$''$}. Due to the low declination of the source and the Yebes latitude ($+$40$^{\circ}$\,31$'$\,29.6$''$), the source was only 5 hours above an elevation of 15$^{\circ}$ per observing session, reaching a maximum elevation of 21$^{\circ}$. We used the new Q band (7\,mm) HEMT receiver that allows broad-band observations in two linear polarizations. The receiver is connected to 16 fast Fourier transform spectrometers (FFTS) with 2.5\,GHz of spectral coverage and 38\,kHz of spectral resolution. This system provides an instantaneous bandwidth of 18\,GHz per polarization in the frequency range between 31.5 and 50\,GHz (F. Tercero et al., in preparation). The intensity scale was calibrated using two absorbers at different temperatures and the atmospheric transmission model (ATM, \citealt{Cernicharo1985,Pardo2001}).

The observational procedure was position switching mode with the reference position located at $\Delta$$\alpha$\,=\,$-$885$''$ and $\Delta$$\delta$\,=\,+290$''$ with respect to G+0.693. The telescope pointing and focus were checked every one or two hours through pseudo-continuum observations towards VX\,Sgr, a red hypergiant star near the target source, with strong SiO $v$\,=\,1 $J$\,=\,1$-$0 (at 43.122\,GHz) maser emission. Pointing errors were always kept within 5$''$ in both axes.

The spectra were measured in units of antenna temperature, T$_{\rm A}^*$, and corrected for atmospheric absorption and spillover losses. At 36\,GHz, system temperatures were in the range 75\,$-$\,180\,K depending on the observing session (with pressure water vapour ranging from 6\,mm to 10\,mm) and source elevation (from 15$^{\circ}$ to 20$^{\circ}$). At this frequency, the aperture efficiency, the conversion factor between flux ($S$) and antenna temperature ($T_{\rm A}^*$), and the half power beam width (HPBW) of the Yebes 40\,m telescope are respectively 0.43, 4.6\,Jy K$^{-1}$, and 48$''$.

In addition, we also used a spectral line survey towards G+0.693 performed with the IRAM 30\,m telescope at Pico Veleta\footnote{IRAM is supported by INSU/CNRS (France), MPG (Germany), and IGN (Spain).} (Spain) and the NRAO\footnote{The National Radio Astronomy Observatories is a facility of the National Science Foundation, operated under a cooperative agreement by Associated Universities, Inc.} 100m Robert C. Byrd Green Bank telescope (GBT) in West Virginia (USA), covering frequencies from 12 to 272\,GHz. The observations were centred at the coordinates \mbox{$\upalpha_{\rm J2000}$\,=\,17$^{\rm h}$\,47$^{\rm m}$\,22$^{\rm s}$}, \mbox{$\updelta_{\rm J2000}$\,=\,$-$28$^{\circ}$\,21$'$\,27$''$}. We refer to \citet{Zeng2018} for more detailed information on the observations.

%%%%%%%%%%%%%%%%%%%%%%%%%%%%%%%%%%%%%%%%%%

\section{Results}
\label{Results}
\subsection{The 1.3 mm continuum map}

The continuum map made with only the SMA data is shown in left panel of Figure \ref{fig:sma-continuum}. The pointing position of G+0.693 ($\alpha$(J2000.0)= 17$^h$ 47$^m$ 22$^s$ and $\delta$(J2000.0)= -28$^{\circ}$ 21$^{\prime}$ 27$^{\prime\prime}$) used in previous IRAM 30\,m surveys \citep[see e.g.][]{Zeng2018} is indicated by a black cross. Consistent with the non-detection of ALMA 3 mm continuum emission in the Sgr B2M and N region recently reported by \citet[][see right panel of Figure \ref{fig:sma-continuum}]{Ginsburg2018}, no clear continuum peak is detected with the SMA at 1.3 mm towards the position of G+0.693. This supports the idea that no massive star formation is taking place within the cloud. To estimate an upper limit to the mass of any possible protostellar core present in the source, we use the following equation:

\begin{equation}
    M = \frac{F_{\nu} d^{2}}{B\rm_v(T\rm_{dust}) \kappa_{\nu}} \frac{M\rm_{gas}}{M\rm_{dust}}
\end{equation}

where $\frac{M\rm_{gas}}{M\rm_{dust}}$ is the gas-to-dust mass ratio, F$_\nu$ is the continuum flux, d is the distance to the target, B$_\nu$(T$_{dust}$) is the Planck function at dust temperature T$_{dust}$, and $\kappa_\nu$ is the dust opacity. We assumed $\frac{M\rm_{gas}}{M\rm_{dust}}$ = 100, d = 8.4 kpc, T$_{dust}$ = 20 K determined from \citet{Rodriguez-fernandez2004,Nagayama2009}, and $\kappa_\nu$ = 0.899 cm$^{2}$g$^{-1}$ \citep[MRN model with thin ice mantles, after 10$^5$ yr of coagulation at densities of 10$^6$ cm$^{-3}$;][]{Ossenkopf1994}\footnote{In the GC, most of the ices have been released into the gas phase because of the shocks, so it is not expected that much of the ices remains in solid state.}. From the 3$\sigma$ rms noise level measured in our SMA 1.3 mm continuum image (15 mJy), we derive an upper limit to the mass of the protostellar envelope of $\leq$2.3 M$\odot$. This value is above the typical lower limit of low-mass Class 0 and \rom{1} envelope masses \citep[>0.5 M$\odot$;][]{Arce2006}, which means that either a low-mass or an intermediate-mass embedded source may exist. To provide further constraints, we carried out the same calculation for the more sensitive ALMA 3 mm continuum from \citet{Ginsburg2018} (see Figure \ref{fig:sma-continuum}). From the 3$\sigma$ rms noise level (2.3 mJy), we derive a protostellar envelope mass $\leq$0.36 M$\odot$, which is below the lower limit of Class 0 envelope masses. This confirms that G+0.693 is quiescent and rules out the possibility that the origin of the molecular complexity found in G+0.693 is due to hidden star formation.

%%%%%%%%%%%%%%%%%%%%%%%%%%%%%%%%%%%%%%%%%%
%%%%%%%%%%%%%%%%%%%%%%%%%%%%%%%%%%%%%%%%%%

\begin{table}
 \centering
 \caption{Parameters of detected molecular lines.}
\begin{adjustbox}{width=\linewidth}
\begin{tabular}{cccc}
\hline
\hline
Molecule & Transition & Rest Frequency & E$\rm_{up}$ \\
& & (GHz) & (K)\\
\hline
\multicolumn{4}{c}{SMA+APEX} \\
\hline
SiO & J=5-4 & 217.1050 & 31.26 \\ 
c-C$_3$H$_2$ & J$\rm_{Ka,Kc}$=6$_{1,6}$-5$_{0,5}$ & 217.8221 & 38.60 \\
c-C$_3$H$_2$ & J$\rm_{Ka,Kc}$=5$_{1,4}$-4$_{2,3}$ & 217.9400 & 35.41 \\
H$_2$CO & J$\rm_{Ka,Kc}$=3$_{0,3}$-2$_{0,2}$ & 218.2222 & 20.96\\
CH$_3$OH & J=4$_{2}$-3$_1$ & 218.4401 & 45.46 \\
H$_2$CO & J$\rm_{Ka,Kc}$=3$_{2,2}$-2$_{2,1}$ & 218.4756 & 68.09\\
HC$_3$N & J=24-23 & 218.3247 & 130.98 \\
H$_2$CO & J$\rm_{Ka,Kc}$=3$_{2,1}$-2$_{2,0}$ & 218.7601 & 68.11 \\
C$^{18}$O & J=2-1 & 219.5603 & 15.80 \\
HNCO & J=10-9 & 219.7983 & 58.02\\
SO & J=6=5 & 219.9494 & 34.98\\
$^{13}$CO & J=2-1 & 220.3986 & 15.86 \\
%CH$_3$OH$^*$ & J=8$_{-1}$-7$_0$ & 229.7587 \\
\hline
\multicolumn{4}{c}{GBT} \\
\hline
CH$_3$OH &  J$\rm_{Ka}$=3$_2$-3$_1$ E & 24.9287 & 36.17 \\
CH$_3$OH &  J$\rm_{Ka}$=4$_2$-4$_1$ E & 24.9335 & 45.46 \\
CH$_3$OH &  J$\rm_{Ka}$=2$_2$-2$_1$ E & 24.9344 & 29.21 \\
CH$_3$OH &  J$\rm_{Ka}$=5$_2$-5$_1$ E & 24.9591 & 57.07 \\
CH$_3$OH &  J$\rm_{Ka}$=6$_2$-6$_1$ E & 25.0181 & 71.00 \\
\hline
\multicolumn{4}{c}{Yebes 40\,m} \\
\hline
CH$_3$OH &  J$\rm_{Ka}$=4$_{-1}$-3$_0$ E & 36.1692 & 28.79\\
CH$_3$OH &  J$\rm_{Ka}$=7$_0$-6$_1$ A & 44.0693 & 64.98\\
\hline
\multicolumn{4}{c}{IRAM 30\,m} \\
\hline
CH$_3$OH &  J$\rm_{Ka}$=5$_{-1}$-4$_0$ E & 84.5211 & 40.39\\
CH$_3$OH &  J$\rm_{Ka}$=8$_0$-7$_1$ A & 95.1693 & 83.54\\
CH$_3$OH &  J$\rm_{Ka}$=6$_{-1}$-5$_0$ E & 132.8907 & 54.31\\
CH$_3$OH &  J$\rm_{Ka}$=8$_{-1}$-7$_0$ E & 229.7587 & 89.10\\
\hline
\hline
\end{tabular}
\end{adjustbox}
\begin{tablenotes}
\item The parameters are taken from the Jet Propulsion Laboratory (JPL) molecular catalogue\footnote{http://spec.jpl.nasa.gov/} \citep{Pickett1998} and the Cologne Database for Molecular Spectroscopy (CDMS)\footnote{http://www.astro.uni-koeln.de/cdms} \citep{Muller2001,Muller2005,Endres2016}.)
\end{tablenotes}
\label{tab:sma-transitions}
\end{table}
%%%%%%%%%%%%%%%%%%%%%%%%%%%%%%%%%%%%%%%%%%
%%%%%%%%%%%%%%%%%%%%%%%%%%%%%%%%%%%%%%%%%%

\subsection{Molecular distributions}

%%%%%%%%%%%%%%%%%%%%%%%%%%%%%%%%%%%%%%%%%%
%%%%%%%%%%%%%%%%%%%%%%%%%%%%%%%%%%%%%%%%%%
\begin{figure*}
    \centering
    \includegraphics[width=\linewidth]{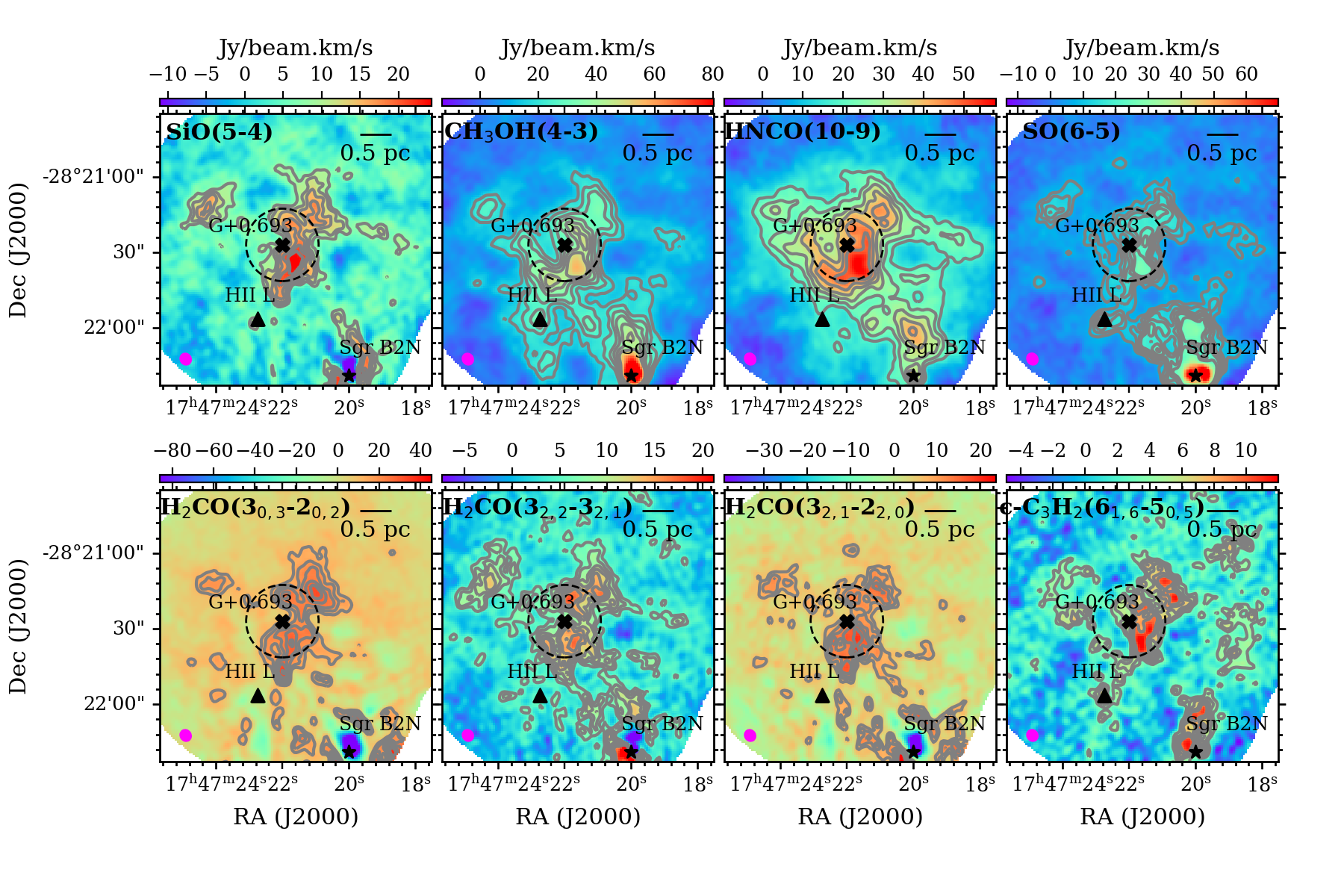}
    \vspace{-9mm}
    \caption{Integrated intensity (moment zeroth) maps of all the detected molecular lines in G+0.693 obtained from the merged SMA+APEX data cubes. The synthesised beam (4$^{\prime\prime}$.4 $\times$ 4$^{\prime\prime}$.0 with a position angle of 33$^{\circ}$.7) is shown as a fucsia ellipse in the bottom left corner. The dashed circle in each panel indicates the beam size ($\sim$29$^{\prime\prime}$) of previous IRAM 30\,m observations towards G+0.693. The grey contour levels represent 30\%-80\%, in the step of 10$\%$, of the molecular peak. The location of Sgr B2N, G+0.693, and H\rom{2} region L are labelled as Figure \ref{fig:sma-continuum}.}
    \label{fig:molecular_distribution}
\end{figure*}
%%%%%%%%%%%%%%%%%%%%%%%%%%%%%%%%%%%%%%%%%%
%%%%%%%%%%%%%%%%%%%%%%%%%%%%%%%%%%%%%%%%%%
 % \vspace*{\floatsep}
%%%%%%%%%%%%%%%%%%%%%%%%%%%%%%%%%%%%%%%%%%
%%%%%%%%%%%%%%%%%%%%%%%%%%%%%%%%%%%%%%%%%%
\begin{figure*}
    \centering
    \includegraphics[width=\linewidth]{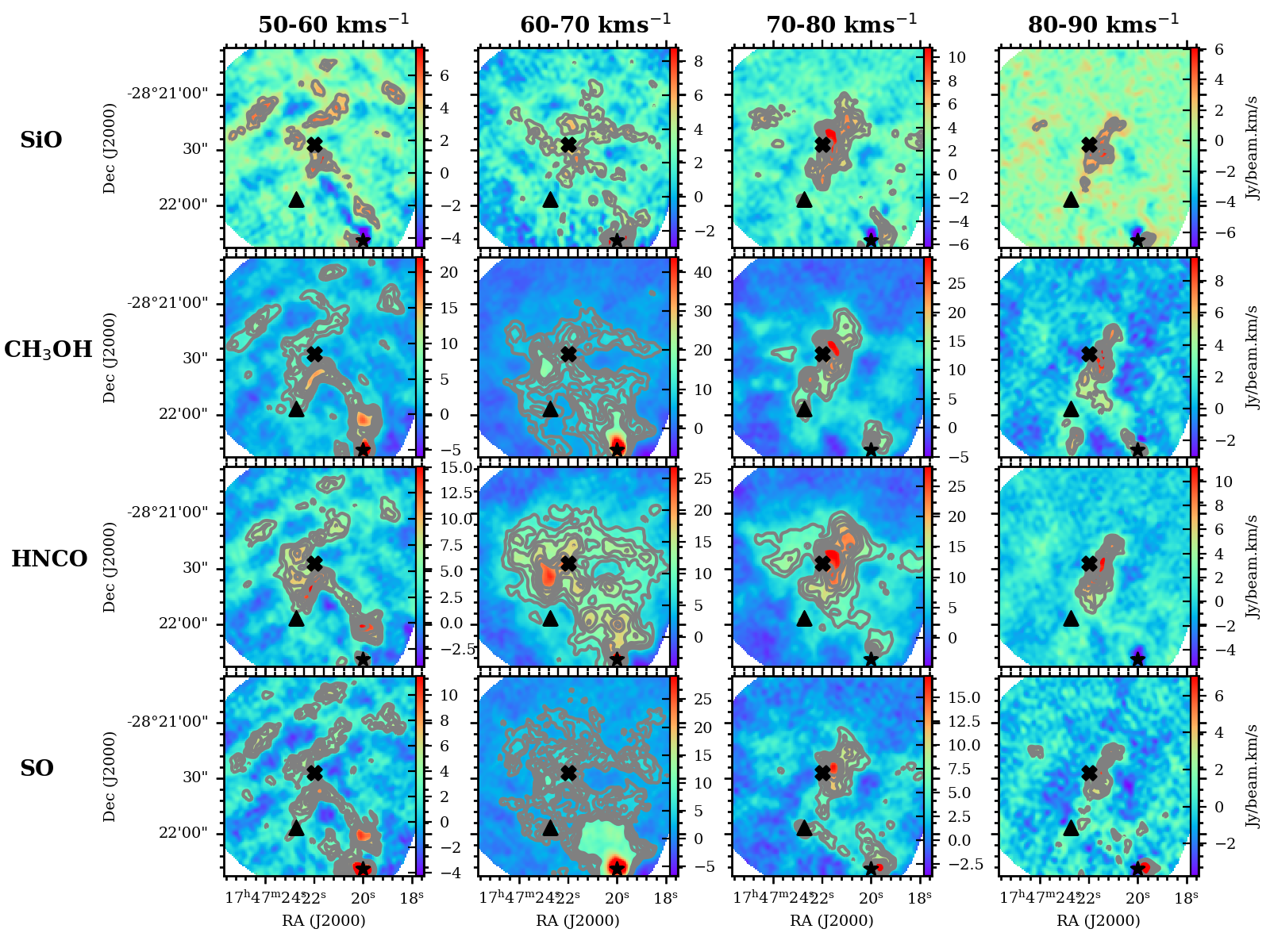}
    \vspace{-4mm}
    \caption{Velocity channel distributions of SiO(5-4), CH$_3$OH(4-3), HNCO(10-9), and SO(6-5) emission at velocity step of 10 km\,s$^{-1}$. In each panel, the contour levels represent 30\%-80\%, in the step of 10$\%$, of the molecular peak.}
    \label{fig:molecular_channel_maps}
\end{figure*}
%%%%%%%%%%%%%%%%%%%%%%%%%%%%%%%%%%%%%%%%%%
%%%%%%%%%%%%%%%%%%%%%%%%%%%%%%%%%%%%%%%%%%

%%%%%%%%%%%%%%%%%%%%%%%%%%%%%%%%%%%%%%%%%%
%%%%%%%%%%%%%%%%%%%%%%%%%%%%%%%%%%%%%%%%%%
\begin{figure*}
    \centering
    \includegraphics[width=\linewidth]{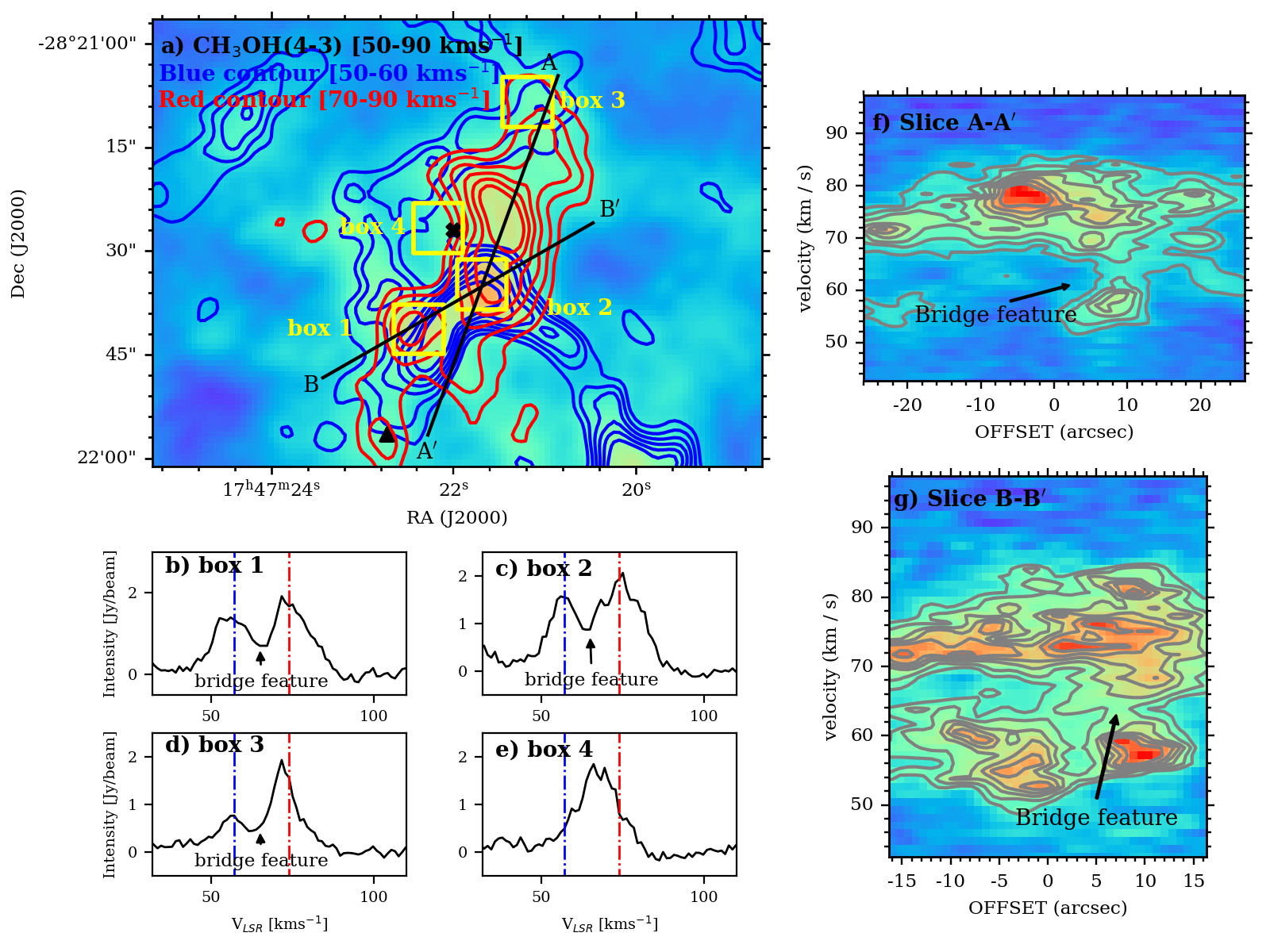}
    \vspace{-0.5cm}
    \caption{(a) Integrated CH$_3$OH(4-3) emission map (colour scale) over velocity range of 50-90 km\,s$^{-1}$ around G+0.693. The blue and red contours indicate the blue-shifted component (50-60 km\,s$^{-1}$) and the red-shifted component (70-90 km\,s$^{-1}$) respectively. The yellow boxes (box 1 to box 4) denote the small areas that are used to obtain an average molecular spectra. The solid lines denoted with A-A$^{\prime}$ and B-B$^{\prime}$ represents the axis where a position-velocity diagram is extracted. (b)-(e) The spectra of CH$_3$OH(4-3) averaged over the corresponding small areas (i.e. box 1 to box 4; see highlighted boxes in (a)). A blue and red vertical broken line indicates the peak velocity of the blue-shifted and the red-shifted components respectively. The bridge features at the intermediate velocity range are annotated by the arrows. (f) A position-velocity diagram along the axis A-A' as shown in Figure 5(a) with a 3-pixel wide slit at P.A.=160$^{\circ}$. (g) A position-velocity diagram along the axis B-B' as shown in (a) with a 3-pixel wide slit at P.A.=120$^{\circ}$.}
    \label{fig:molecular_pv}
\end{figure*}
%%%%%%%%%%%%%%%%%%%%%%%%%%%%%%%%%%%%%%%%%%
%%%%%%%%%%%%%%%%%%%%%%%%%%%%%%%%%%%%%%%%%%

As summarised in Table \ref{tab:sma-transitions}, 12 molecular lines were detected in G+0.693 by the combination of SMA and APEX observations. Spectra of the SMA-only full 4\,GHz lower sideband are presented in Figure \ref{fig:full-spectra}. We note that for most of the molecular species analysed here, we only have one detected transition each detected in our data set. This does not allow us to accurately determine the abundances of these molecules towards this source. This aspect of the analysis will be covered in a forthcoming paper using higher sensitivity single-dish surveys carried out towards G+0.693 with the IRAM 30\,m and the Yebes 40\,m telescopes. Although $^{13}$CO (J=2-1) is detected towards G+0.693, it is excluded from the analysis as its emission is ubiquitous in the region and its spatial distribution cannot be distinguished between G+0.693 and Sgr B2N. Note that for c-C$_3$H$_2$ (5$_{1,4}$-4$_{2,3}$), HC$_3$N, and C$^{18}$O, their emissions are not strong enough to show significant kinematic feature in the emission map, hence they will also be excluded from the analysis. Among the rest of the detected lines, we show in Figure \ref{fig:molecular_distribution} their velocity-integrated intensity, represented by the zeroth-moment maps. The maps were made by integrating the molecular emission over a velocity range between $\sim$50 and $\sim$90 km\,s$^{-1}$. The overall distribution of molecular gas is extended and generally show an elongated morphology which stretches along the north-south direction. In particular, strong molecular emission tend to distribute in an arch shape that surrounds the west side of G+0.693. Moreover, a molecular condensation is clearly seen in the same off-position, $\sim$5$^{\prime\prime}$ ($\sim$0.2 pc) south-west away from the pointing position of G+0.693. Given that the beam size of the IRAM 30\,m observations is $\sim$29$^{\prime\prime}$ (indicated by a dashed circle in Figure \ref{fig:molecular_distribution}), this shift in the peak position of molecular emissions is conceivable and may imply the presence of smaller scale structure that has never been observed previously. Furthermore, the observed condensation might be a density enhancement produced by shocked gas which might represent the precursor of massive star-forming regions. The relatively strong emission of well-known shock tracers such as SiO, CH$_3$OH, HNCO, and SO is shown to be tracing the shocked materials around G+0.693. This further supports the proposed mechanism that wealth of COMs observed towards G+0.693 is likely due to widespread shocks sputtering off grain mantles, releasing molecular species into the gas phase \citep{Martin-pintado1997,Requena-torres2006}.

 In the following, we will mainly focus on the shock tracers: HNCO, SiO, CH$_3$OH, and SO, detected in our dataset to investigate the kinematics of G+0.693. In Figure \ref{fig:molecular_channel_maps}, we present the integrated intensity maps of SiO, CH$_3$OH, HNCO, and SO at four different velocity ranges (i.e. 50-60, 60-70, 70-80, and 80-90 km\,s$^{-1}$). At 50-60 km\,s$^{-1}$, molecular gas is distributed mainly on the east side in regard to the position of G+0.693. On the other hand, the prominent emission peak seen in Figure \ref{fig:molecular_distribution} distinctly appears on the west side at 70-90 km\,s$^{-1}$. The entangled gas distribution present at 60-70 km\,s$^{-1}$ may indicate that the lower and higher velocity components are merged at this intermediate velocity range. We hereafter refer the component at 50-60 km\,s$^{-1}$ 'blue-shifted component' and the component at 70-90 km\,s$^{-1}$ 'red-shifted component'. In Figure \ref{fig:molecular_pv}(a), the blue-shifted component and red-shifted component (blue and red contour respectively) are superimposed on the CH$_3$OH integrated intensity map (colour scale). We also include the same figure for HNCO, SO, and SiO emission in the Appendix as supplementary materials. It is interesting to note that the northern and the southern part of these two components are spatially overlapping while the middle region is apart from each other. Since we have combined the SMA observation with the APEX data, this middle region is not caused by missing flux of the interferometric data. To further examine the two velocity components, we display the spectra towards our selected four small areas highlighted by the respective box in Figure \ref{fig:molecular_pv}(a). Note that boxes 1 and 3 are chosen at the southern and the northern region where two clouds are overlapped; box 2 is chosen at the molecular peak observed in zeroth-moment maps (see Figure \ref{fig:molecular_distribution}); box 4 is chosen at an area where two clouds are apart from each other. Each spectrum is obtained by averaging each area (see Figure \ref{fig:molecular_pv}(b) - (e)). For the spectrum obtained from areas where the blue-shifted component and the red-shifted component is overlapped (i.e. box 1,2, and 3), the two velocity components centred at $\sim$57 and $\sim$75 km\,s$^{-1}$ are presented along with an almost flattened profile between two velocity peaks. The latter can be interpreted as the so-called 'bridge feature', an observational signature of cloud-cloud collision, at the intermediate velocity range which can also be seen in a position-velocity diagram (more details in Discussions). Indeed, the position-velocity diagrams of CH$_3$OH (see Figure \ref{fig:molecular_pv} (f) and (g)) depict consistently the bridge feature connecting the two clouds in velocity space. This feature is also noticeable in the positive-velocity diagrams of other shock tracers (see Figures \ref{fig:molecular_pv_HNCO}-\ref{fig:molecular_pv_SO} in Appendix).

\subsection{Methanol maser lines in G+0.693}
\label{maser}

%%%%%%%%%%%%%%%%%%%%%%%%%%%%%%%%%%%%%%%%%%
%%%%%%%%%%%%%%%%%%%%%%%%%%%%%%%%%%%%%%%%%%
\begin{figure*}
    \centering
    \includegraphics[width=\linewidth]{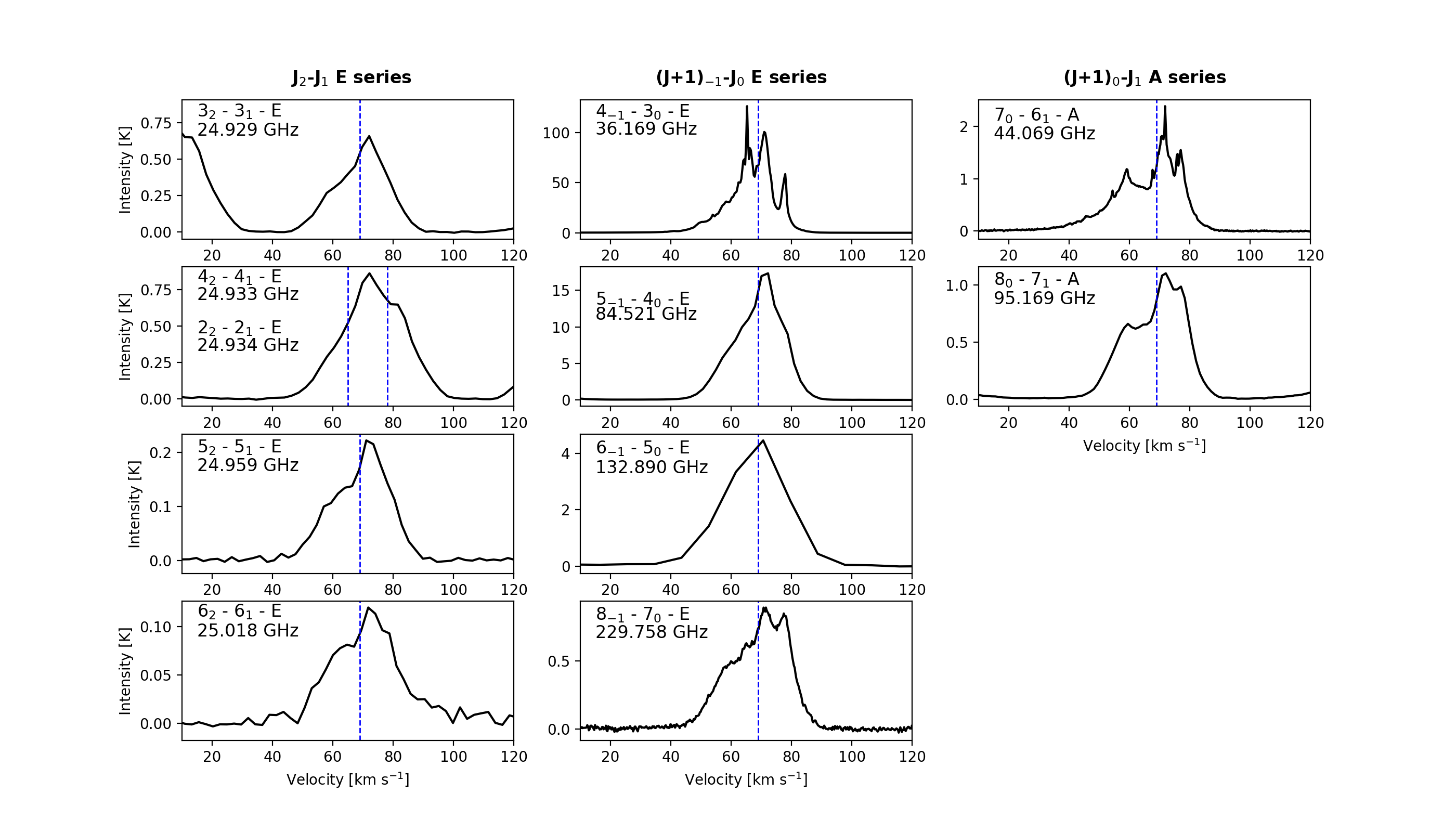}
    \vspace{-0.7cm}
    \caption{Spectra of the class \rom{1} methanol masers detected towards G+0.693. The blue dashed line denote the central radial velocity of G+0.693. The corresponding transition and frequency is indicated at the top left corner in each panel.}
    \label{fig:CH3OH_masers}
\end{figure*}
%%%%%%%%%%%%%%%%%%%%%%%%%%%%%%%%%%%%%%%%%%
%%%%%%%%%%%%%%%%%%%%%%%%%%%%%%%%%%%%%%%%%%

%%%%%%%%%%%%%%%%%%%%%%%%%%%%%%%%%%%%%%%%%%
%%%%%%%%%%%%%%%%%%%%%%%%%%%%%%%%%%%%%%%%%%

\begin{figure*}
    \begin{minipage}[b]{0.49\textwidth}
    \includegraphics[width=\textwidth]{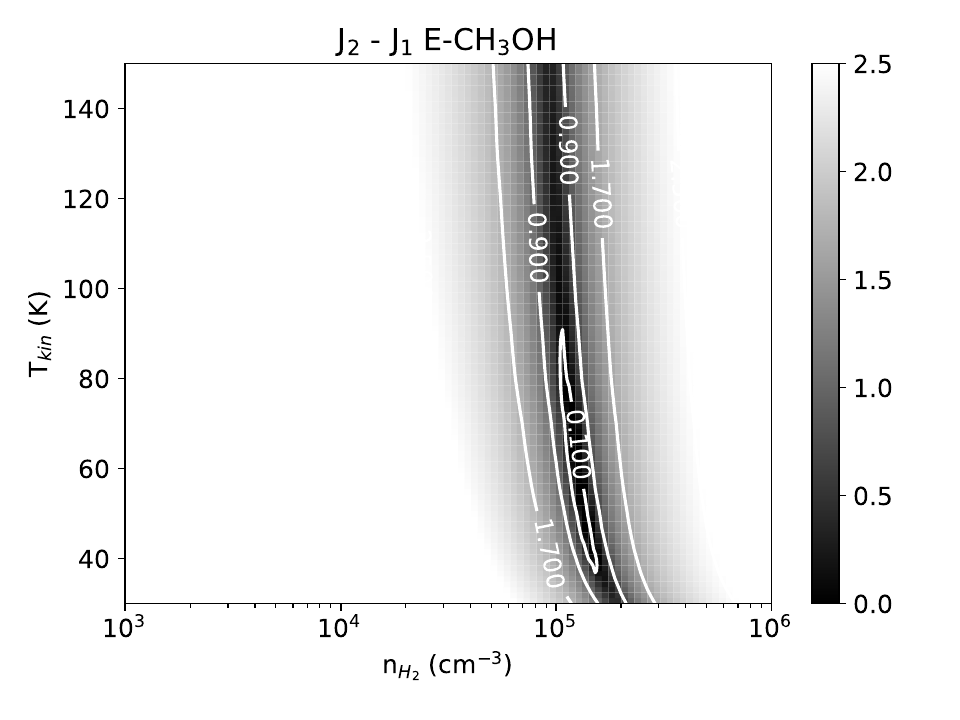}
  \end{minipage}
  \hfill
  \begin{minipage}[b]{0.49\textwidth}
    \includegraphics[width=\textwidth]{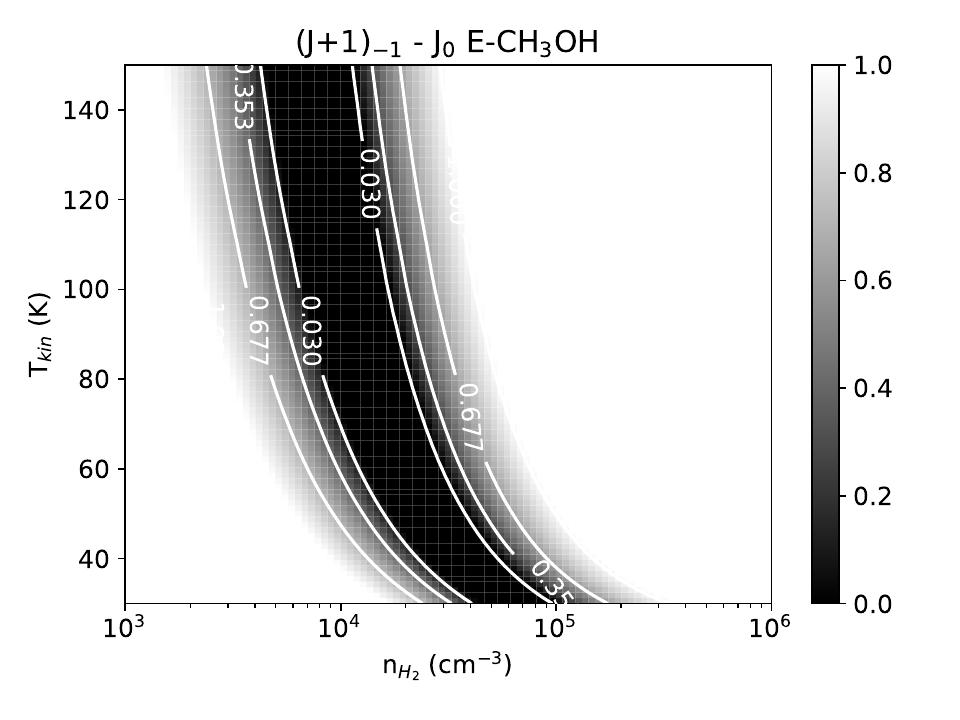}
  \end{minipage}

%  \begin{subfigure}[b]{0.49\linewidth}
%    \includegraphics[width=\linewidth]{e-ch3oh_25.pdf}
%  \end{subfigure}
  %
%  \begin{subfigure}[b]{0.49\linewidth}
%    \includegraphics[width=\linewidth]{e-ch3oh.pdf}
%  \end{subfigure}
\caption{Log of $\chi ^2$ fit results from \text{radex} analysis for J$_2$ - J$_1$ E series and (J+1)$_{-1}$ - J$_0$ E series of class I methanol masers for a wide range of kinetic temperatures and H$_2$ densities. The darker regions show a lower $\chi ^2$ which indicates better fits than lighter regions. }
\label{fig:radex-fit}
\end{figure*}

%\begin{figure*}
%    \centering
%    \includegraphics[width=\textwidth]{maser_radex_plot.png}
%    \vspace{-0.5cm}
%    \caption{RADEX modelling results for the J$_2$ - J$_1$ E series and the (J+1)$_{-1}$ - J$_0$ E %series of class I methanol masers. Left panel: peak intensity of each transition at various %n$\rm_{H_2}$ with fixed T$\rm_{kin}$. Middle panel: peak intensity of each transition at various %T$\rm_{kin}$ with fixed n$\rm_{H_2}$. Right panel: best fit model. Black cross denotes the observed %peak intensity.}
%    \label{fig:radex-fit}
%\end{figure*}
%
%%%%%%%%%%%%%%%%%%%%%%%%%%%%%%%%%%%%%%%%%%
%%%%%%%%%%%%%%%%%%%%%%%%%%%%%%%%%%%%%%%%%%

Class \rom{1} methanol masers are considered as good tracers of outflow activity in star-forming regions. It is remarkable that a rather strong class \rom{1} methanol maser is detected in G+0.693 where no star formation is taken place. Therefore, the class \rom{1} methanol maser is unlikely to be associated with outflow activity. Indeed, unlike outflows, the methanol maser emission at 36\,GHz is extended over several parsecs across the SgrB2 complex \citep[see][]{Liechti1996,Jones2011}, which suggests that the physical conditions in the large-scale shocks give rise to the class \rom{1} methanol maser in the Sgr B2 region. Using the Yebes 40\,m telescope, we have detected the two well-known class \rom{1} methanol maser lines 4$_{-1}$ - 3$_0$ E and 7$_0$ - 6$_1$ A at 36.2 and 44.1\,GHz respectively towards G+0.693. Together with the spectral line surveys obtained with GBT and IRAM 30\,m telescope, 11 class \rom{1} methanol maser lines in total are identified towards G+0.693 (see Table \ref{tab:sma-transitions}). They can be categorised into three families: the J$_2$ - J$_1$ E series, the (J+1)$_{-1}$ - J$_0$ E series, and the (J+1)$_0$ - J$_1$ A series \citep{Leurini2016}. All the line profiles of the detected class \rom{1} methanol maser lines are presented in Figure \ref{fig:CH3OH_masers}. Owing to the larger beam sizes of GBT, IRAM 30\,m, and Yebes 40\,m observations, the presence of two velocity components are not as clear as that shown in the SMA+APEX data, but they seem to appear at the same corresponding velocity range, most notable in (J+1)$_0$ - J$_1$ A series. 

The 36 and 44\,GHz methanol masers have already been reported in the Sgr B2 region \citep{Liechti1996,Jones2011}. The emission of these two lines are shown to have different spatial distribution, in particular the 36\,GHz line shows its strongest emission towards the position of G+0.693 whilst the peak of 44\,GHz emission appears near Sgr B2N and Sgr B2M \citep[see Figure 4 in][]{Jones2011}. For the line profile obtained in this study, multiple narrow velocity features are present in both 36 and 44\,GHz lines which are likely associated to inhomogeneities in the shocked gas. They also consist of broad ($\sim$10-20 km\,s$^{-1}$) components as well as narrow (<5 km\,s$^{-1}$) spike-like ones. However the two line profiles seem to be complementary in velocity, the brightest peak of 36\,GHz line is offset with respect to the 44\,GHz line. This possibly indicates that the masers are arising from different regions of the cloud. Furthermore, the 36\,GHz emission appears much more prominent than the 44\,GHz line towards G+0.693. This is rather unusual since majority of the sources with both 36 and 44\,GHz lines detected are enhanced in 44\,GHz \citep{Voronkov2014}.

By using the multiple CH$_3$OH transitions covered in our dataset, we attempt to constrain the physical properties of the gas that is experiencing masering amplification by using the non-LTE radiative transfer code \textsc{radex} \citep{Vandetak2007} with collision rates from \citet{Rabli2010}. To run \textsc{radex}, we assume a uniform spherical geometry and a cosmic microwave background radiation temperature of T = 2.73 K. Given that the molecular emission is extended, the beam-filling factor is assumed to be unity. The CH$_3$OH linewidth and column density were fixed to $\delta$v = 20 km\,s$^{-1}$ which is the derived median values of CH$_3$OH linewidth from five different transitions detected in G+0.693 \citep{Requena-torres2008}; and N$_{CH_3OH}$ = 2.1$\times$10$^{16}$ cm$^{-2}$ which is calculated from the derived CH$_3$OH abundance (5$\times$10$^{-7}$) with respect to H$_2$ by adopting their derived H$_2$ column density = 4.1$\times$10$^{22}$ cm$^{-2}$ \citep{Requena-torres2008}. The model grids encompass H$_2$ densities in the range of n$\rm_{H_2}$ = 10$^3$ - 10$^6$ cm$^{-3}$ and kinetic temperatures T$_{kin}$ range between 30 - 150 K. The line intensities generated by \textsc{radex} for all transitions can then be compared to the measured line intensities. A reduced $\chi^2$ fitting was conducted for each model to determine the best fit. Our results are displayed in Figure \ref{fig:radex-fit}. Although both J$_2$ - J$_1$ and (J+1)$_{-1}$ - J$_0$ series belong to E-CH$_3$OH species, \text{radex} does not converge when all the transitions are considered simultaneously. Indeed, different excitation conditions are required to reproduce the CH$_3$OH maser emissions from the two series. For J$_2$ - J$_1$ E series, we obtain the minimum reduced $\chi ^2$ for a density $\sim$ 1.0$\times$10$^{5}$ cm$^{-3}$ and temperature in a range of 40 - 90 K. While for (J+1)$_{-1}$ - J$_0$ E series, the density is well fit between 10$^{4}$ and 10$^{5}$ cm$^{-3}$ but the temperature is poorly constrained as low $\chi ^2$ values are found across the temperature range. With only two transitions detected for (J+1)$_0$-J$_1$ A series, the models cannot provide constraints on density and temperature. In general, our results agree with the finding of \citet{Leurini2016} in which the class \rom{1} masers in the 25\,GHz series i.e. the J$_2$ - J$_1$ E series mase at higher densities than other lines. They are also consistent with the physical conditions of typical Galactic Centre molecular clouds with little or no star formation activity. The difference in excitation conditions support the idea that these maser series likely arise from different parts of the cloud. It is noteworthy that the methanol emission at 218\,GHz (5$_2$-4$_1$) which is an analogue of the 25\,GHz series \citep{Voronkov2012,Hunter2014} does not show inversion population in our calculation. This means that the 218\,GHz line is quasi-thermal and it reflects the gas distribution in G+0.693.

%%%%%%%%%%%%%%%%%%%%%%%%%%%%%%%%%%%%%%%%%%
%%%%%%%%%%%%%%%%%%%%%%%%%%%%%%%%%%%%%%%%%%
%\begin{table}
% \centering
% \caption{Derived parameters of the detected CH$_3$OH (J+1)$_2$-J$_1$ E and (J+1)$_{-1}$ - J$_0$ E %series for the best fit RADEX models.}
%\begin{adjustbox}{width=\linewidth}
%\begin{tabular}{cccccc}
%\hline
%\hline
%Transition & Rest Frequency & T$\rm_{ex}$ & $\tau$ & T$\rm_R$ & Observed value \\
% & (GHz) & (K) & & (K) & (K)\\
%\hline
%J$\rm _{Ka}$=3$_2$-3$_1$ E & 24.9287 & -6.3 & -0.07 & 0.63 & 0.65\\
%J$\rm _{Ka}$=4$_2$-4$_1$ E & 24.9335 & -5.1 & -0.06 & 0.45 & 0.43\\
%J$\rm _{Ka}$=2$_2$-2$_1$ E & 24.9344 & -5.9 & -0.07 & 0.63 & 0.60\\
%J$\rm _{Ka}$=5$_2$-5$_1$ E & 24.9591 & -5.3 & -0.03 & 0.22 & 0.22\\
%J$\rm _{Ka}$=6$_2$-6$_1$ E & 25.0181 & -5.1 & -0.01 & 0.09 & 0.12\\
%\hline
%J$\rm _{Ka}$=4$_{-1}$-3$_0$ E & 36.1692 & -1.1 & -3.25 & 102.2 & 100.50\\
%J$\rm_{Ka}$=5$_{-1}$-4$_0$ E & 84.5211 & -2.7 & -1.04 & 11.67 & 15.00 \\
%J$\rm_{Ka}$=6$_{-1}$-5$_0$ E & 132.8907 & -4.4 & -0.22 & 2.22 & 4.41 \\
%J$\rm_{Ka}$=8$_{-1}$-7$_0$ E & 229.7587 & -7.4 & -0.01 & 0.20 & 0.89 \\
%\hline
%J$\rm _{Ka}$=7$_0$-6$_1$ A & 44.0693 & - & - & - & 1.85\\
%J$\rm_{Ka}$=8$_0$-7$_1$ A & 95.1693 & - & - & - & 1.10 \\
%\hline
%\hline
%\end{tabular}
%\end{adjustbox}
%\begin{tablenotes}
%\item  
%\end{tablenotes}
%\label{tab:maser-radex}
%\end{table}
%%%%%%%%%%%%%%%%%%%%%%%%%%%%%%%%%%%%%%%%%%%
%%%%%%%%%%%%%%%%%%%%%%%%%%%%%%%%%%%%%%%%%%

%%%%%%%%%%%%%%%%%%%%%%%%%%%%%%%%%%%%%%%%%%

\section{Discussions}
\label{Discussions}

%%%%%%%%%%%%%%%%%%%%%%%%%%%%%%%%%%%%%%%%%%
%%%%%%%%%%%%%%%%%%%%%%%%%%%%%%%%%%%%%%%%%%
\begin{figure*}
    \centering
    \includegraphics[width=0.7\textwidth]{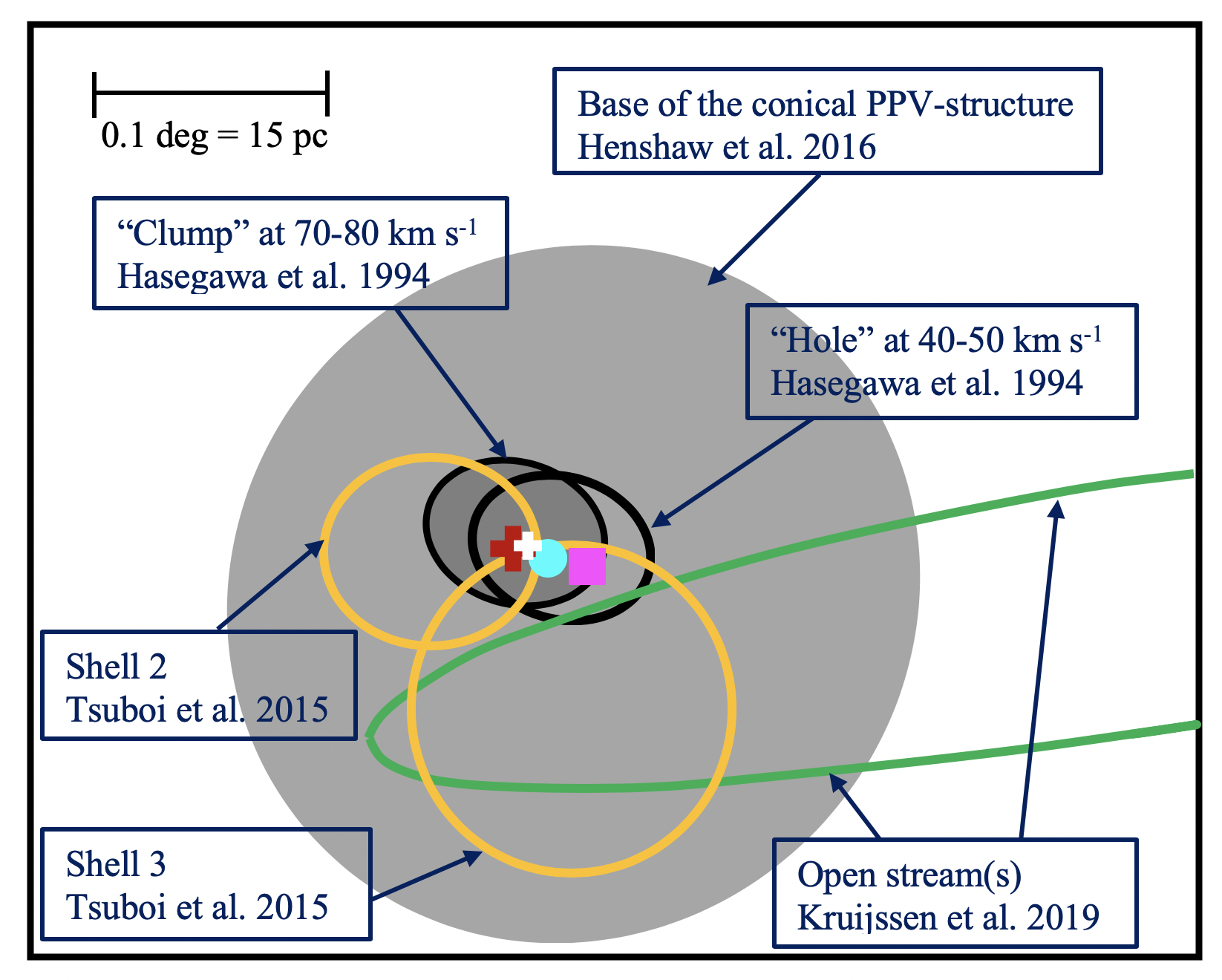}
    %\vspace{-3cm}
    \caption{A sketch of different observed kinematic features towards G+0.693 region. G+0.693, Sgr B2N, and Sgr B2M are denoted by red cross, cyan circle, and magenta square respectively. The white cross and grey circle indicate respectively the tip and the base of the conical PPV-structure reported in \citet{Henshaw2016}. The black and shaded ellipse refer to the 'hole' and the 'clump' features observed in \citet{Hasegawa1994}. The orange circles denote the expanding shell 2 and 3 in \citet{Tsuboi2015}. The green arcs denote the open stream(s) proposed by \citet{Kruijssen2019}.}
    \label{fig:sketch}
\end{figure*}
%%%%%%%%%%%%%%%%%%%%%%%%%%%%%%%%%%%%%%%%%%
%%%%%%%%%%%%%%%%%%%%%%%%%%%%%%%%%%%%%%%%%%

\subsection{Comparisons with observations in Sgr B2 region}
Towards the Sgr B2 molecular cloud complex, \citet{Hasegawa1994} discovered the characteristic kinematic features of a 'hole' and a 'clump' region in $^{13}$CO distribution (see the sketch in Figure \ref{fig:sketch}. The 'hole' refers to a shell-like structure with an inner cavity roughly 5$^{\prime}$ in size and that appears for the emission integrated between velocities of 40 - 50 km\,s$^{-1}$. The 'clump' refers to a prominent emission peak that shows up at the same position with similar morphology by integrating over the velocity range of 70 - 80 km\,s$^{-1}$. As illustrated in Figure \ref{fig:sketch}, G+0.693 is located within the 'hole' and 'clump' region. Consistently, little molecular emission are present in our study at the velocity range of 40-50 km\,s$^{-1}$ while the molecular emission peak is found at 70-80 km\,s$^{-1}$. However, our dataset does not cover a large enough area to reveal the morphological complementarity between the 'hole' and the 'clump' features as reported in \citet{Hasegawa1994} and \citet{Sato2000}. In order to explain the existence of these features, a cloud-cloud collision scenario has been put forward \citep{Hasegawa1994}. The 'hole' at velocities between 40 and 50 km\,s$^{-1}$ is presumably created by a dense clump with velocities between 70 and 80 km\,s$^{-1}$ that is plowing into a more extended molecular cloud, resulting in a cloud-cloud collision with an inclination to the northeast with respect to the line of sight, producing strong shocks in the interface region. This hypothesis is further supported by \citet{Sato2000}. In such case, G+0.693 is possibly situated near the northern interface where the interaction would be expected to be stronger. Indeed, our observations seems to draw the same picture where the red-shifted components at velocities between 70 and 80 km\,s$^{-1}$ (corresponding to the 'clump') interacts with the blue-shifted components at velocities between 50 and 60 km\,s$^{-1}$, causing the deficient emission (the 'hole') observed at velocities between 40 and 50 km\,s$^{-1}$. And the shock tracers detected towards G+0.693 likely trace the shock associated with this cloud-cloud collision. 

The gas kinematics of the Sgr B2 region have recently been revisited by \citet{Henshaw2016} in which a conical structure is observed in position-position-velocity (PPV) space with HNCO. Figure \ref{fig:sketch} only displays the base and the tip of this conical structure. In comparison to the 'hole' and the 'clump' features, the PPV-structure of Sgr B2 is continuous, i.e. there is no discontinuity between the velocity of the 'hole' and the emission peak. As a consequence, the impression of a hole feature can be seen by integrating over different velocity ranges except for the emission peak due to the conical structure, but not only at velocity of 40-50 km\,s$^{-1}$ as previously observed. In addition, \citet{Henshaw2016} found that the emission peak lies between 60 and 70 km\,s$^{-1}$, does not correspond to the 'clump' at 70-80 km\,s$^{-1}$. And the molecular emission between 70 and 100 km\,s$^{-1}$ is considered to be related to an extended high-velocity feature present at the north-east region beyond Sgr B2 complex. 

This conical structure however, can have various interpretations. One could argue that the cone shape is in fact the result of a high-velocity cloud that is punching through a cloud at lower velocity, supporting the cloud-cloud collision scenario proposed by \citet{Hasegawa1994}. The appearance of a hole can be easily created due to the conical profile and/or self-absorption of molecular line profile. Indeed, the $^{13}$CO emission used in \citet{Hasegawa1994} is known to be affected by self-absorption \citep{Sato2000}. Furthermore, the tip of the conical structure almost overlaps with the position of G+0.693 (see Figure \ref{fig:sketch}) which implies the same idea that G+0.693 is expected to be affected by the shock from the nearby cloud-cloud collision. On the contrary, such conical structure is speculated to be closely linked to the orbital dynamics of the gas in the CMZ \citep{Kruijssen2019}. In the models described by \citet{Kruijssen2015,Henshaw2016,Kruijssen2019}, the gas in the inner CMZ (R<120 pc) is represented by open steam(s) following an eccentric orbit. Sgr B2 resides close to the apocentre where the orbit curves off, causing the superposition of clouds along the line of sight (see Figure \ref{fig:sketch}). The conical structure observed is suggested to be naturally created in this manner without invoking the event of cloud-cloud collision. 

In spite of these, several shell-like features are identified by \citet[][labelled as Shells 1-6]{Tsuboi2015} in the Sgr B2 complex in SiO and H$^{13}$CO$^+$. The detailed parameters of these shells are given in Table 4 of their paper. In Figure \ref{fig:sketch}, it shows that G+0.693 is close to the rim of Shell 2 and Shell 3, possibly residing at the region where the two shells intersect. Although both Shell 2 and Shell 3 are proposed to be expanding shells, the former is considered to be originated from successive supernovae in the CMZ while the latter is postulated to be the result of an ongoing cloud-cloud collision with the time-scale of $\sim$10$^{5}$ yr \citep[see also][on expanding shell in NH$_3$]{Martin-Pintado1999}.

%%%%%%%%%%%%%%%%%%%%%%%%%%%%%%%%%%%%%%%%%%
%%%%%%%%%%%%%%%%%%%%%%%%%%%%%%%%%%%%%%%%%%
%\begin{figure}
%    \centering
%    \includegraphics[width=\linewidth]{SMA_hole_clump.png}
%    \vspace{-3cm}
%    \caption{Integrated emission map of SiO(5-4), CH$_3$OH(4-3), HNCO(10-9), and SO(6-5) at velocity range of 40-50 km\,s$^{-1}$. The name of molecule is given in the top-right corner of each panel. In each panel, white and grey contour levels represent 30\%-80\%, in the step of 10$\%$, of the molecular peak at 40-50 km\,s$^{-1}$ and at 70-80 km\,s$^{-1}$ respectively.}
%    \label{fig:molecular_hole_clump}
%\end{figure}
%%%%%%%%%%%%%%%%%%%%%%%%%%%%%%%%%%%%%%%%%%
%%%%%%%%%%%%%%%%%%%%%%%%%%%%%%%%%%%%%%%%%%

%%%%%%%%%%%%%%%%%%%%%%%%%%%%%%%%%%%%%%%%%%
%%%%%%%%%%%%%%%%%%%%%%%%%%%%%%%%%%%%%%%%%%
%\begin{figure}
%    \centering
%    \includegraphics[width=\linewidth]{SMA_Tsuboi_shells.png}
%    \vspace{-1cm}
%    \caption{Integrated emission map of SiO(5-4) at velocity range of 50-90 km\,s$^{-1}$. The blue and red contours indicate the blue-shifted component (50-60 km\,s$^{-1}$) and the red-shifted component (70-90 km\,s$^{-1}$) respectively. Shell 2 and Shell 3 identified in \citet{Tsuboi2015} are depicted in green and brown circles respectively. Note the thickness of the circle does not represent the true thickness of the shell. The arrows denote the shell is expanding.}
%    \label{fig:molecular_Tsuboi_shells}
%\end{figure}
%%%%%%%%%%%%%%%%%%%%%%%%%%%%%%%%%%%%%%%%%%
%%%%%%%%%%%%%%%%%%%%%%%%%%%%%%%%%%%%%%%%%%

\subsection{Scenario of an early cloud-cloud collision?}

Our present observations have revealed the following pieces of evidence about the gas kinematics towards G+0.693. 

\begin{itemize}
    \item From the velocity channel maps, two velocity components 50-60 and 70-90 km\,s$^{-1}$ are identified. 
    \item The distribution of two components overlaps towards the southern and northern part while the middle part is spatially separated. 
    \item In the position-velocity diagram, the red-shifted component and blue-shifted component are connected by a bridge feature. 
\end{itemize}

At a glance, Figure \ref{fig:molecular_pv}(a) suggests that two clouds are likely interacting, possibly colliding against each other. Hence in the following, we will interpret these observational features by making comparisons with the common observational and theoretical signatures characteristic of the cloud-cloud collision scenario. As one of the outcomes of cloud-cloud collision, complementary distribution between two colliding clouds at distinct velocities is expected \citep{Habe1992,Anathpindika2010, Takahira2014, Torii2017, Fukui2018}. Such complementary distribution is presented particularly between colliding clouds with different sizes in which the smaller cloud creates a cavity in the larger cloud. From the model prediction of collisions between two spheres of different sizes, Figure 6 in \citet{Takahira2014} presented the snapshot of surface density plots of the collision model at different relative velocities. The complementary distribution that is accounted for by the cavity is determined by the size as well as the travel distance of the smaller cloud since the initiation of the collision. In addition, depending on the angle of the relative motion to the line of sight ($\theta$), the smaller cloud can be coincident ($\theta$=0) or displaced ($\theta \neq$0) with the intensity becoming depressed in the molecular distribution at the velocity of the larger cloud \citep{Fukui2018}. 

In the case of G+0.693, the two velocity components do not show clear complementary distribution but overlap with each other for most part of the cloud. Comparing our results with the collision models of \citet{Takahira2014}, this may indicate the collision have just been initiated and the cavity has not yet been created by the smaller cloud. Several observations towards young high-mass star-forming regions that are proposed to be at early stage of cloud-cloud collisions also showed less or non-existent complementary gas distribution between the colliding clouds \citep[e.g.][]{Fukui2016,Fukui2018b,Hayashi2020}

Another observational support of cloud-cloud collision is the presence of the bridge feature. It represents the shocked interface layer in-between the two clouds and it manifests itself as a velocity component connecting the two clouds at the intermediate velocity in a position-velocity diagram. \citet{Haworth2015b,Haworth2015a} presented synthetic position-velocity diagrams based on hydrodynamical simulations from \citet{Takahira2014} and \citet{Shima2016}. Considering the velocity separation of $\sim$20 km\,s$^{-1}$ between two clouds identified in G+0.693, the bridge feature revealed in position velocity diagram (Figure \ref{fig:molecular_pv} (f) and (g)) seems to be consistent with snapshot of these diagrams taken between 0.4 and 2 Myr after the onset of the collision \citep[see Figure 7 in][]{Haworth2015b}. This agrees with the absence of complementary distribution in our observations as the collision has just begun and the bridge feature has started to form. Nevertheless, the strong enhancement of SiO emission towards G+0.693 may serve as evidence for supersonic shock waves that are generated at the beginning of a collision \citep[e.g.][]{Hasegawa1994,Tsuboi2015}.

The detection of bright class \rom{1} methanol masers towards G+0.693 also strengthens the scenario of cloud-cloud collision. They usually act as tracers of interstellar shocked gas. Particularly, studies have found class \rom{1} methanol masers to be excited in the shocked regions associated with molecular outflows \citep[e.g.][]{Kurtz2004,Voronkov2006, Cyganowski2009}, interaction between supernova remnants and molecular clouds \citep[e.g.][]{Sjouwerman2010,Frail2011,Pihlstrom2011,Pihlstrom2014}, interaction between expanding H \rom{2} regions and the ambient molecular environment \citep[e.g.][]{Voronkov2010a,Voronkov2012} and cloud-cloud collisions \citep[e.g.][]{Sobolev1992,Salii2002}. In the CMZ, the interaction between enhanced cosmic rays and molecular gas can also be responsible for class \rom{1} methanol masers \citep{Yusef-Zadeh2013}. 

Opposite to star forming regions with protestellar outflows, where the 44\,GHz line is usually found to be stronger than the other methanol maser transitions \citep[e.g.][]{Pratap2008,Voronkov2014}, the 36\,GHz line appears to be more than an order of magnitude brighter than the 44\,GHz transition towards G+0.693. Besides, no studies so far have reported in G+0.693 accompanying emission from 6.7\,GHz class \rom{2} masers, 22\,GHz water masers and other transitions that are typically found in star-forming regions \citep{Ladeyschikov2019,Lu2019b}. All this suggests that molecular outflows and expanding H \rom{2} regions can hardly be responsible for the methanol masers in G+0.693. Although the 36\,GHz line is observed to be more intense than the 44\,GHz line in regions shocked by supernova remnant-molecular cloud interactions \citep[e.g.][]{Pihlstrom2011,Pihlstrom2014}, they are expected to be accompanied by OH masers at 1720 MHz \citep[e.g.][]{Yusef-Zadeh2003,Frail2011}. towards G+0.693, the absence of OH masers as well as other observational signatures of supernova remnant-molecular cloud interaction summarised by \citet{Slane2015} indicate that such interaction is unlikely the mechanism to produce methanol masers detected in this study. 

From our available observational data, one remarkable result to note is that the strongest integrated intensity over velocity between 70 and 80 km\,s$^{-1}$ (see column 3 in Figure \ref{fig:molecular_channel_maps} of shock tracers SiO, CH$_3$OH, HNCO, and SO superimpose spatially with the location where the most intense 36 and 84\,GHz methanol emission is observed by \citet{Liechti1996}, \citet{Jones2008,Jones2011} (roughly at $\alpha$(J2000.0)= 17$^h$ 47$^m$ 21$^s$ and $\delta$(J2000.0)= -28$^{\circ}$ 21$^{\prime}$ 23$^{\prime\prime}$). This is specifically the position located right above box 2 where the blue-shifted component interacts with the red-shifted component in Figure \ref{fig:molecular_pv}. The enhanced abundance of SiO and HNCO observed towards G+0.693 is suggested to be produced by large-scale low-velocity shocks present in the region \citep{Martin-pintado1997,Minh2006,Martin2008}. From this perspective, shocks generated by the collisions between two clouds are favoured in our case to provide the excitation responsible for the class \rom{1} methanol maser. 

A similar case is reported in the extragalactic source NGC 253 \citep{Ellingsen2017}. Other than detecting 36\,GHz line about two orders of magnitude higher than the 44\,GHz line, the integrated intensity of SiO is shown to be coincident with that of the 36\,GHz methanol emission. The authors proposed that the methanol maser emission originates in a region with a significant rate of cloud-cloud collisions. The alike interpretation of class \rom{1} methanol masers between G+0.693 and NGC 253 is consistent with previous studies that show a striking similarity in chemical abundances and excitation conditions between the two regions \citep{Martin2006,Armijos2015}.

\subsection{Origin of the rich chemistry in G+0.693}

G+0.693 is a molecular cloud in the Galactic Centre that exhibits a very rich chemistry, comparable to that observed in star-forming regions in the Galactic disc. However, the SMA and ALMA continuum maps show that G+0.693 does not seem to be associated with any thermal continuum source and the dust temperature ($\leq$20 K) is considerably lower than the gas temperature (30 - 150 K). Both of these imply that stellar feedback is likely not responsible for the high abundances of molecular species such as HNCO or CH3OH measured in this position \citep[e.g.][]{Requena-torres2006,Zeng2018}. A low excitation temperature of $\leq$20 K has been found for numerous COMs detected towards G+0.693 \citep[e.g.][]{Requena-torres2008,Zeng2018}, which can be explained by the low gas density (10$^4$-10$^5$ cm$^{-3}$) of the cloud that allows the sub-thermal excitation of COMs.

On the other hand, as a consequence of a cloud-cloud collision, shocks are expected to sputter efficiently the icy grain mantles ejecting molecules formed on grains into the gas phase. Such mechanical process does not only explain the extended distribution of different molecular emission revealed in this study but also the high gas-phase abundance in this region. A large condensation of gas is found to coincide with the position where two velocity components likely interact in G+0.693. In other words, the gas condensation may be formed by the cloud-cloud collision which has enhanced the gas density of the region and the induced shocks would activate the rich chemistry in G+0.693. The \textsc{radex} analysis of J$_2$ - J$_1$ E series of methanol masers is found to trace higher gas density than (J+1)$_{-1}$ - J$_0$ E series meaning that the collision giving rise to the methanol maser J$_2$ - J$_1$ E series is in the higher post-shock density. However spatial distribution of J$_2$ - J$_1$ E series is required to confirm the coincidence between the region where this maser originates from within the region where the collision occur. Furthermore, it is natural to assume the physical and chemical properties of the surrounding gas are also modified by the shock driven by this cloud-cloud collision. This is supported by the presence of another molecular condensation that is observed in shock tracers such as SiO, HNCO, SO and CH$_3$OH by integrating intensity over velocities between 70 and 80 km\,s$^{-1}$. This position also coincides with that reported for the strong emission of (J+1)$_{-1}$ - J$_0$ E methanol maser. From the \textsc{radex} modelling results, we find that the gas kinetic temperature is hard to be constrained which could be due to the E$\rm_{up}$ of the detected transitions being lower than the actual T$\rm_{kin}$. It may alternatively indicate this region is not at a constant temperature and this is not surprising in a shocked region where temperature varies rapidly with time and space.

The CMZ in the Galactic Centre is rather a complex environment where many energetic phenomenon can play a role in the chemistry of the molecular gas. For instance, X-rays and the enhanced cosmic rays ionisation rates have been invoked to account for some of the abundant COMs detected in G+0.693 \citep{Zeng2018}. All the arguments above emphasise the suggestion that shocks induced by the cloud-cloud collision is likely the most important process responsible for the high level of chemical complexity observed towards G+0.693.

%%%%%%%%%%%%%%%%%%%%%%%%%%%%%%%%%%%%%%%%%%
%%%%%%%%%%%%%%%%%%%%%%%%%%%%%%%%%%%%%%%%%%
%\begin{figure}
%    \centering
%    \includegraphics[width=0.7\linewidth]{Yebes_CH3OH_maser_zoom.png}
%    %\vspace{-2cm}
%    \caption{}
%    \label{fig:CH3OH_maser_zoom}
%\end{figure}
%%%%%%%%%%%%%%%%%%%%%%%%%%%%%%%%%%%%%%%%%%
%%%%%%%%%%%%%%%%%%%%%%%%%%%%%%%%%%%%%%%%%%

\section{Conclusions}
\label{Conclusions}

For the first time, we have studied the small-scale morphology and kinematics of G+0.693 by using interferometric data in combination with single-dish observations. The major outcomes of the paper are as follow:

\begin{itemize}
    \item No clear continuum peak is detected in the 1.3 mm continuum map obtained with the SMA, supporting the quiescent nature of G+0.693. This is consistent with previous studies carried out with ALMA at 3 mm \citep{Ginsburg2016}.
    \item From the SMA spectral line observations and complementary APEX sing-dish data, we found that the general molecular gas distribution towards G+0.693 is extended and elongated in the north-south direction. A molecular condensation is revealed in an offset position from the pointing position of G+0.693 which implies the existence of substructure that has not been uncovered previously.
    \item Two molecular components appear at the velocity ranges of 50-60 and 70-90 km\,s$^{-1}$ as identified from the molecular gas emission of shock tracers such as HNCO, SiO, CH$_3$OH and HNCO. Their integrated emission map and position-velocity diagrams depict observational characteristic of a cloud-cloud collision that has just been initiated. 
    \item A total of 11 transitions from three different series of class \rom{1} methanol masers are detected towards G+0.693 in our GBT, Yebes 40\,m and IRAM 30\,m single-dish data. This type of masers are associated with shock interactions, supporting the idea of a large-scale cloud-cloud collision. We modelled the multiple transitions of J$_2$ - J$_1$ E series and (J+1)$_{-1}$ - J$_0$ E series with non-LTE calculations to provide constraints on the physical conditions in G+0.693. We obtained different excitation conditions for two series which may indicate they arise from different regions in G+0.693. The analysis also provides a strong constraint on the gas density which is between 10$^4$ and 10$^5$ cm$^{-3}$ but not on the gas temperature. The large range of the gas temperature obtained from the models may be explained by the idea that the class \rom{1} methanol masers originate from a shocked region which is presumably affected by the event of a cloud-cloud collision.
    \item The characteristics of the class \rom{1} methanol masers studied in G+0.693 appear to share remarkable similarities with the class \rom{1} masers detected in the extragalactic source NGC 253, which has also been proposed to be experiencing cloud-cloud collisions in its nuclear starburst. 
\end{itemize}

In summary, our results are consistent with the proposed idea that the chemistry of G+0.693 is dominated by low-velocity shocks which are likely originated from the occurrence of a cloud-cloud collision. If this holds, it is expected that such collision would enhance the gas density to levels at which star formation may proceed.

\section*{Acknowledgements}
We wish to thank the anonymous referee for his/her very useful comments that helped to improve this article. We would like to thank Adam Ginsburg (University of Florida) for providing the ALMA image. Based on observations with the 40-m radio telescope of the National Geographic Institute of Spain (IGN) at Yebes Observatory (project number 20A008). Yebes Observatory acknowledges the ERC for funding support under grant ERC-2013-Syg-610256-NANOCOSMOS. S. Z acknowledges support through a Principal's studentship funded by Queen Mary University of London, the visiting student program funded by Harvard-Smithsonian Center for Astrophysics. I.J.-S. and J.M.-P. have received partial support from the Spanish FEDER under project number ESP2017-86582-C4-1-R. V.M.R. has received funding from the European Union's Horizon 2020 research and innovation programme under the Marie Sk\l{}odowska-Curie grant agreement No 664931. X.L. was financially supported by JSPS KAKENHI grants No.\ 18K13589 \& 20K14528

%%%%%%%%%%%%%%%%%%%%%%%%%%%%%%%%%%%%%%%%%%

\section*{Data Availability}
The data underlying this article will be shared on reasonable request to the corresponding author.



\bibliographystyle{mnras}
\bibliography{references.bib} % if your bibtex file is called example.bib

\begin{thebibliography}{}
\makeatletter
\relax
\def\mn@urlcharsother{\let\do\@makeother \do\$\do\&\do\#\do\^\do\_\do\%\do\~}
\def\mn@doi{\begingroup\mn@urlcharsother \@ifnextchar [ {\mn@doi@}
  {\mn@doi@[]}}
\def\mn@doi@[#1]#2{\def\@tempa{#1}\ifx\@tempa\@empty \href
  {http://dx.doi.org/#2} {doi:#2}\else \href {http://dx.doi.org/#2} {#1}\fi
  \endgroup}
\def\mn@eprint#1#2{\mn@eprint@#1:#2::\@nil}
\def\mn@eprint@arXiv#1{\href {http://arxiv.org/abs/#1} {{\tt arXiv:#1}}}
\def\mn@eprint@dblp#1{\href {http://dblp.uni-trier.de/rec/bibtex/#1.xml}
  {dblp:#1}}
\def\mn@eprint@#1:#2:#3:#4\@nil{\def\@tempa {#1}\def\@tempb {#2}\def\@tempc
  {#3}\ifx \@tempc \@empty \let \@tempc \@tempb \let \@tempb \@tempa \fi \ifx
  \@tempb \@empty \def\@tempb {arXiv}\fi \@ifundefined
  {mn@eprint@\@tempb}{\@tempb:\@tempc}{\expandafter \expandafter \csname
  mn@eprint@\@tempb\endcsname \expandafter{\@tempc}}}

\bibitem[\protect\citeauthoryear{{Anathpindika}}{{Anathpindika}}{2010}]{Anathpindika2010}
{Anathpindika} S.~V.,  2010, \mn@doi [\mnras]
  {10.1111/j.1365-2966.2010.16541.x}, \href
  {https://ui.adsabs.harvard.edu/abs/2010MNRAS.405.1431A} {405, 1431}

\bibitem[\protect\citeauthoryear{{Arce} \& {Sargent}}{{Arce} \&
  {Sargent}}{2006}]{Arce2006}
{Arce} H.~G.,  {Sargent} A.~I.,  2006, \mn@doi [\apj] {10.1086/505104}, \href
  {https://ui.adsabs.harvard.edu/abs/2006ApJ...646.1070A} {646, 1070}

\bibitem[\protect\citeauthoryear{{Armijos-Abenda{\~n}o}, {Mart{\'\i}n-Pintado},
  {Requena-Torres}, {Mart{\'\i}n}  \&
  {Rodr{\'\i}guez-Franco}}{{Armijos-Abenda{\~n}o} et~al.}{2015}]{Armijos2015}
{Armijos-Abenda{\~n}o} J.,  {Mart{\'\i}n-Pintado} J.,  {Requena-Torres} M.~A.,
  {Mart{\'\i}n} S.,   {Rodr{\'\i}guez-Franco} A.,  2015, \mn@doi [\mnras]
  {10.1093/mnras/stu2271}, \href
  {https://ui.adsabs.harvard.edu/abs/2015MNRAS.446.3842A} {446, 3842}

\bibitem[\protect\citeauthoryear{{Bally}, {Stark}, {Wilson}  \&
  {Henkel}}{{Bally} et~al.}{1987a}]{Bally1987}
{Bally} J.,  {Stark} A.~A.,  {Wilson} R.~W.,   {Henkel} C.,  1987a, \mn@doi
  [\apjs] {10.1086/191217}, \href
  {http://adsabs.harvard.edu/abs/1987ApJS...65...13B} {65, 13}

\bibitem[\protect\citeauthoryear{{Bally}, {Stark}  \& {Wilson}}{{Bally}
  et~al.}{1987b}]{Bally1987a}
{Bally} J.,  {Stark} A.~A.,   {Wilson} R.~W.,  1987b, in {Peimbert} M.,
  {Jugaku} J.,  eds,  IAU Symposium Vol. 115, Star Forming Regions. p.~550

\bibitem[\protect\citeauthoryear{{Battersby} et~al.,}{{Battersby}
  et~al.}{2017}]{Battersby2017}
{Battersby} C.,  et~al., 2017, in {Crocker} R.~M.,  {Longmore} S.~N.,
  {Bicknell} G.~V.,  eds,  IAU Symposium Vol. 322, The Multi-Messenger
  Astrophysics of the Galactic Centre. pp 90--94 (\mn@eprint {arXiv}
  {1610.05805}), \mn@doi{10.1017/S1743921316012266}

\bibitem[\protect\citeauthoryear{{Battersby} et~al.,}{{Battersby}
  et~al.}{2020}]{Battersby2020}
{Battersby} C.,  et~al., 2020, arXiv e-prints, \href
  {https://ui.adsabs.harvard.edu/abs/2020arXiv200705023B} {p. arXiv:2007.05023}

\bibitem[\protect\citeauthoryear{{Cernicharo}}{{Cernicharo}}{1985}]{Cernicharo1985}
{Cernicharo} J.,  1985, Internal IRAM Report (Granada: IRAM)

\bibitem[\protect\citeauthoryear{{Cyganowski}, {Brogan}, {Hunter}  \&
  {Churchwell}}{{Cyganowski} et~al.}{2009}]{Cyganowski2009}
{Cyganowski} C.~J.,  {Brogan} C.~L.,  {Hunter} T.~R.,   {Churchwell} E.,  2009,
  \mn@doi [\apj] {10.1088/0004-637X/702/2/1615}, \href
  {https://ui.adsabs.harvard.edu/abs/2009ApJ...702.1615C} {702, 1615}

\bibitem[\protect\citeauthoryear{{De Vicente}, {Mart{\'\i}n-Pintado}, {Neri}
  \& {Colom}}{{De Vicente} et~al.}{2000}]{deVicente2000}
{De Vicente} P.,  {Mart{\'\i}n-Pintado} J.,  {Neri} R.,   {Colom} P.,  2000,
  \aap, \href {https://ui.adsabs.harvard.edu/abs/2000A&A...361.1058D} {361,
  1058}

\bibitem[\protect\citeauthoryear{{Ellingsen}, {Chen}, {Breen}  \&
  {Qiao}}{{Ellingsen} et~al.}{2017}]{Ellingsen2017}
{Ellingsen} S.~P.,  {Chen} X.,  {Breen} S.~L.,   {Qiao} H.~H.,  2017, \mn@doi
  [\mnras] {10.1093/mnras/stx2076}, \href
  {https://ui.adsabs.harvard.edu/abs/2017MNRAS.472..604E} {472, 604}

\bibitem[\protect\citeauthoryear{{Endres}, {Schlemmer}, {Schilke}, {Stutzki}
  \& {M{\"u}ller}}{{Endres} et~al.}{2016}]{Endres2016}
{Endres} C.~P.,  {Schlemmer} S.,  {Schilke} P.,  {Stutzki} J.,   {M{\"u}ller}
  H.~S.~P.,  2016, \mn@doi [Journal of Molecular Spectroscopy]
  {10.1016/j.jms.2016.03.005}, \href
  {https://ui.adsabs.harvard.edu/abs/2016JMoSp.327...95E} {327, 95}

\bibitem[\protect\citeauthoryear{{Frail}}{{Frail}}{2011}]{Frail2011}
{Frail} D.~A.,  2011, \memsai, \href
  {https://ui.adsabs.harvard.edu/abs/2011MmSAI..82..703F} {82, 703}

\bibitem[\protect\citeauthoryear{{Fukui} et~al.,}{{Fukui}
  et~al.}{2016}]{Fukui2016}
{Fukui} Y.,  et~al., 2016, \mn@doi [\apj] {10.3847/0004-637X/820/1/26}, \href
  {https://ui.adsabs.harvard.edu/abs/2016ApJ...820...26F} {820, 26}

\bibitem[\protect\citeauthoryear{{Fukui} et~al.,}{{Fukui}
  et~al.}{2018a}]{Fukui2018b}
{Fukui} Y.,  et~al., 2018a, \mn@doi [\pasj] {10.1093/pasj/psy017}, \href
  {https://ui.adsabs.harvard.edu/abs/2018PASJ...70S..41F} {70, S41}

\bibitem[\protect\citeauthoryear{{Fukui} et~al.,}{{Fukui}
  et~al.}{2018b}]{Fukui2018}
{Fukui} Y.,  et~al., 2018b, \mn@doi [\apj] {10.3847/1538-4357/aac217}, \href
  {https://ui.adsabs.harvard.edu/abs/2018ApJ...859..166F} {859, 166}

\bibitem[\protect\citeauthoryear{{Ginsburg} et~al.,}{{Ginsburg}
  et~al.}{2016}]{Ginsburg2016}
{Ginsburg} A.,  et~al., 2016, \mn@doi [\aap] {10.1051/0004-6361/201526100},
  \href {http://adsabs.harvard.edu/abs/2016A\%26A...586A..50G} {586, A50}

\bibitem[\protect\citeauthoryear{{Ginsburg} et~al.,}{{Ginsburg}
  et~al.}{2018}]{Ginsburg2018}
{Ginsburg} A.,  et~al., 2018, \mn@doi [\apj] {10.3847/1538-4357/aaa6d4}, \href
  {http://adsabs.harvard.edu/abs/2018ApJ...853..171G} {853, 171}

\bibitem[\protect\citeauthoryear{{Goto}, {Indriolo}, {Geballe}  \&
  {Usuda}}{{Goto} et~al.}{2013}]{Goto2013}
{Goto} M.,  {Indriolo} N.,  {Geballe} T.~R.,   {Usuda} T.,  2013, \mn@doi
  [Journal of Physical Chemistry A] {10.1021/jp400017s}, \href
  {https://ui.adsabs.harvard.edu/abs/2013JPCA..117.9919G} {117, 9919}

\bibitem[\protect\citeauthoryear{{Guesten} \& {Henkel}}{{Guesten} \&
  {Henkel}}{1983}]{Guesten1983}
{Guesten} R.,  {Henkel} C.,  1983, \aap, \href
  {https://ui.adsabs.harvard.edu/abs/1983A&A...125..136G} {125, 136}

\bibitem[\protect\citeauthoryear{{Guesten}, {Walmsley}, {Ungerechts}  \&
  {Churchwell}}{{Guesten} et~al.}{1985}]{Guesten1985}
{Guesten} R.,  {Walmsley} C.~M.,  {Ungerechts} H.,   {Churchwell} E.,  1985,
  \aap, \href {http://adsabs.harvard.edu/abs/1985A\%26A...142..381G} {142, 381}

\bibitem[\protect\citeauthoryear{{Habe} \& {Ohta}}{{Habe} \&
  {Ohta}}{1992}]{Habe1992}
{Habe} A.,  {Ohta} K.,  1992, \pasj, \href
  {https://ui.adsabs.harvard.edu/abs/1992PASJ...44..203H} {44, 203}

\bibitem[\protect\citeauthoryear{{Hasegawa}, {Sato}, {Whiteoak}  \&
  {Miyawaki}}{{Hasegawa} et~al.}{1994}]{Hasegawa1994}
{Hasegawa} T.,  {Sato} F.,  {Whiteoak} J.~B.,   {Miyawaki} R.,  1994, \mn@doi
  [\apjl] {10.1086/187417}, \href
  {http://adsabs.harvard.edu/abs/1994ApJ...429L..77H} {429, L77}

\bibitem[\protect\citeauthoryear{{Hasegawa}, {Arai}, {Yamaguchi}  \&
  {Sato}}{{Hasegawa} et~al.}{2008}]{Hasegawa2008}
{Hasegawa} T.,  {Arai} T.,  {Yamaguchi} N.,   {Sato} F.,  2008, \mn@doi [\apss]
  {10.1007/s10509-007-9523-7}, \href
  {https://ui.adsabs.harvard.edu/abs/2008Ap&SS.313...91H} {313, 91}

\bibitem[\protect\citeauthoryear{{Haworth} et~al.,}{{Haworth}
  et~al.}{2015a}]{Haworth2015b}
{Haworth} T.~J.,  et~al., 2015a, \mn@doi [\mnras] {10.1093/mnras/stv639}, \href
  {https://ui.adsabs.harvard.edu/abs/2015MNRAS.450...10H} {450, 10}

\bibitem[\protect\citeauthoryear{{Haworth}, {Shima}, {Tasker}, {Fukui},
  {Torii}, {Dale}, {Takahira}  \& {Habe}}{{Haworth}
  et~al.}{2015b}]{Haworth2015a}
{Haworth} T.~J.,  {Shima} K.,  {Tasker} E.~J.,  {Fukui} Y.,  {Torii} K.,
  {Dale} J.~E.,  {Takahira} K.,   {Habe} A.,  2015b, \mn@doi [\mnras]
  {10.1093/mnras/stv2068}, \href
  {https://ui.adsabs.harvard.edu/abs/2015MNRAS.454.1634H} {454, 1634}

\bibitem[\protect\citeauthoryear{{Hayashi} et~al.,}{{Hayashi}
  et~al.}{2020}]{Hayashi2020}
{Hayashi} K.,  et~al., 2020, arXiv e-prints, \href
  {https://ui.adsabs.harvard.edu/abs/2020arXiv200507933H} {p. arXiv:2005.07933}

\bibitem[\protect\citeauthoryear{{Henshaw} et~al.,}{{Henshaw}
  et~al.}{2016}]{Henshaw2016}
{Henshaw} J.~D.,  et~al., 2016, \mn@doi [\mnras] {10.1093/mnras/stw121}, \href
  {http://adsabs.harvard.edu/abs/2016MNRAS.457.2675H} {457, 2675}

\bibitem[\protect\citeauthoryear{{Huettemeister}, {Wilson}, {Bania}  \&
  {Martin-Pintado}}{{Huettemeister} et~al.}{1993}]{Huettemeister1993}
{Huettemeister} S.,  {Wilson} T.~L.,  {Bania} T.~M.,   {Martin-Pintado} J.,
  1993, \aap, \href {http://adsabs.harvard.edu/abs/1993A\%26A...280..255H}
  {280, 255}

\bibitem[\protect\citeauthoryear{{Hunter}, {Brogan}, {Cyganowski}  \&
  {Young}}{{Hunter} et~al.}{2014}]{Hunter2014}
{Hunter} T.~R.,  {Brogan} C.~L.,  {Cyganowski} C.~J.,   {Young} K.~H.,  2014,
  \mn@doi [\apj] {10.1088/0004-637X/788/2/187}, \href
  {https://ui.adsabs.harvard.edu/abs/2014ApJ...788..187H} {788, 187}

\bibitem[\protect\citeauthoryear{{Jim{\'e}nez-Serra}, {Mart{\'\i}n-Pintado},
  {Rodr{\'\i}guez-Franco}, {Chandler}, {Comito}  \&
  {Schilke}}{{Jim{\'e}nez-Serra} et~al.}{2007}]{Jimenez-Serra2007}
{Jim{\'e}nez-Serra} I.,  {Mart{\'\i}n-Pintado} J.,  {Rodr{\'\i}guez-Franco} A.,
   {Chandler} C.,  {Comito} C.,   {Schilke} P.,  2007, \mn@doi [\apjl]
  {10.1086/519005}, \href
  {https://ui.adsabs.harvard.edu/abs/2007ApJ...661L.187J} {661, L187}

\bibitem[\protect\citeauthoryear{{Jim{\'e}nez-Serra}, {Caselli},
  {Mart{\'\i}n-Pintado}  \& {Hartquist}}{{Jim{\'e}nez-Serra}
  et~al.}{2008}]{Jimenez-Serra2008}
{Jim{\'e}nez-Serra} I.,  {Caselli} P.,  {Mart{\'\i}n-Pintado} J.,   {Hartquist}
  T.~W.,  2008, \mn@doi [\aap] {10.1051/0004-6361:20078054}, \href
  {https://ui.adsabs.harvard.edu/abs/2008A&A...482..549J} {482, 549}

\bibitem[\protect\citeauthoryear{{Jim{\'e}nez-Serra}, {Mart{\'\i}n-Pintado},
  {Caselli}, {Mart{\'\i}n}, {Rodr{\'\i}guez-Franco}, {Chandler}  \&
  {Winters}}{{Jim{\'e}nez-Serra} et~al.}{2009}]{Jimenez-Serra2009}
{Jim{\'e}nez-Serra} I.,  {Mart{\'\i}n-Pintado} J.,  {Caselli} P.,
  {Mart{\'\i}n} S.,  {Rodr{\'\i}guez-Franco} A.,  {Chandler} C.,   {Winters}
  J.~M.,  2009, \mn@doi [\apjl] {10.1088/0004-637X/703/2/L157}, \href
  {https://ui.adsabs.harvard.edu/abs/2009ApJ...703L.157J} {703, L157}

\bibitem[\protect\citeauthoryear{{Jimenez-Serra} et~al.,}{{Jimenez-Serra}
  et~al.}{2020}]{Jimenez-Serra2020}
{Jimenez-Serra} I.,  et~al., 2020, arXiv e-prints, \href
  {https://ui.adsabs.harvard.edu/abs/2020arXiv200407834J} {p. arXiv:2004.07834}

\bibitem[\protect\citeauthoryear{{Jones} et~al.,}{{Jones}
  et~al.}{2008}]{Jones2008}
{Jones} P.~A.,  et~al., 2008, \mn@doi [\mnras]
  {10.1111/j.1365-2966.2008.13009.x}, \href
  {http://adsabs.harvard.edu/abs/2008MNRAS.386..117J} {386, 117}

\bibitem[\protect\citeauthoryear{{Jones}, {Burton}, {Tothill}  \&
  {Cunningham}}{{Jones} et~al.}{2011}]{Jones2011}
{Jones} P.~A.,  {Burton} M.~G.,  {Tothill} N.~F.~H.,   {Cunningham} M.~R.,
  2011, \mn@doi [\mnras] {10.1111/j.1365-2966.2010.17849.x}, \href
  {https://ui.adsabs.harvard.edu/abs/2011MNRAS.411.2293J} {411, 2293}

\bibitem[\protect\citeauthoryear{{Kauffmann}, {Pillai}, {Zhang}, {Menten},
  {Goldsmith}, {Lu}, {Guzm{\'a}n}  \& {Schmiedeke}}{{Kauffmann}
  et~al.}{2017}]{Kauffmann2017}
{Kauffmann} J.,  {Pillai} T.,  {Zhang} Q.,  {Menten} K.~M.,  {Goldsmith} P.~F.,
   {Lu} X.,  {Guzm{\'a}n} A.~E.,   {Schmiedeke} A.,  2017, \mn@doi [\aap]
  {10.1051/0004-6361/201628089}, \href
  {https://ui.adsabs.harvard.edu/abs/2017A&A...603A..90K} {603, A90}

\bibitem[\protect\citeauthoryear{{Koyama}, {Awaki}, {Kunieda}, {Takano}  \&
  {Tawara}}{{Koyama} et~al.}{1989}]{Koyama1989}
{Koyama} K.,  {Awaki} H.,  {Kunieda} H.,  {Takano} S.,   {Tawara} Y.,  1989,
  \mn@doi [\nat] {10.1038/339603a0}, \href
  {https://ui.adsabs.harvard.edu/abs/1989Natur.339..603K} {339, 603}

\bibitem[\protect\citeauthoryear{{Krieger} et~al.,}{{Krieger}
  et~al.}{2017}]{Krieger2017}
{Krieger} N.,  et~al., 2017, \mn@doi [\apj] {10.3847/1538-4357/aa951c}, \href
  {http://adsabs.harvard.edu/abs/2017ApJ...850...77K} {850, 77}

\bibitem[\protect\citeauthoryear{{Kruijssen}, {Dale}  \&
  {Longmore}}{{Kruijssen} et~al.}{2015}]{Kruijssen2015}
{Kruijssen} J.~M.~D.,  {Dale} J.~E.,   {Longmore} S.~N.,  2015, \mn@doi
  [\mnras] {10.1093/mnras/stu2526}, \href
  {https://ui.adsabs.harvard.edu/abs/2015MNRAS.447.1059K} {447, 1059}

\bibitem[\protect\citeauthoryear{{Kruijssen} et~al.,}{{Kruijssen}
  et~al.}{2019}]{Kruijssen2019}
{Kruijssen} J.~M.~D.,  et~al., 2019, \mn@doi [\mnras] {10.1093/mnras/stz381},
  \href {https://ui.adsabs.harvard.edu/abs/2019MNRAS.484.5734K} {484, 5734}

\bibitem[\protect\citeauthoryear{{Kurtz}, {Hofner}  \& {{\'A}lvarez}}{{Kurtz}
  et~al.}{2004}]{Kurtz2004}
{Kurtz} S.,  {Hofner} P.,   {{\'A}lvarez} C.~V.,  2004, \mn@doi [\apjs]
  {10.1086/423956}, \href
  {https://ui.adsabs.harvard.edu/abs/2004ApJS..155..149K} {155, 149}

\bibitem[\protect\citeauthoryear{{Ladeyschikov}, {Bayandina}  \&
  {Sobolev}}{{Ladeyschikov} et~al.}{2019}]{Ladeyschikov2019}
{Ladeyschikov} D.~A.,  {Bayandina} O.~S.,   {Sobolev} A.~M.,  2019, \mn@doi
  [\aj] {10.3847/1538-3881/ab4b4c}, \href
  {https://ui.adsabs.harvard.edu/abs/2019AJ....158..233L} {158, 233}

\bibitem[\protect\citeauthoryear{{Leurini}, {Menten}  \& {Walmsley}}{{Leurini}
  et~al.}{2016}]{Leurini2016}
{Leurini} S.,  {Menten} K.~M.,   {Walmsley} C.~M.,  2016, \mn@doi [\aap]
  {10.1051/0004-6361/201527974}, \href
  {https://ui.adsabs.harvard.edu/abs/2016A&A...592A..31L} {592, A31}

\bibitem[\protect\citeauthoryear{{Liechti} \& {Wilson}}{{Liechti} \&
  {Wilson}}{1996}]{Liechti1996}
{Liechti} S.,  {Wilson} T.~L.,  1996, \aap, \href
  {https://ui.adsabs.harvard.edu/abs/1996A&A...314..615L} {314, 615}

\bibitem[\protect\citeauthoryear{{Longmore} et~al.,}{{Longmore}
  et~al.}{2013}]{Longmore2013}
{Longmore} S.~N.,  et~al., 2013, \mn@doi [\mnras] {10.1093/mnras/sts376}, \href
  {https://ui.adsabs.harvard.edu/abs/2013MNRAS.429..987L} {429, 987}

\bibitem[\protect\citeauthoryear{{Lu} et~al.,}{{Lu} et~al.}{2019a}]{Lu2019b}
{Lu} X.,  et~al., 2019a, \mn@doi [\apjs] {10.3847/1538-4365/ab4258}, \href
  {https://ui.adsabs.harvard.edu/abs/2019ApJS..244...35L} {244, 35}

\bibitem[\protect\citeauthoryear{{Lu} et~al.,}{{Lu} et~al.}{2019b}]{Lu2019a}
{Lu} X.,  et~al., 2019b, \mn@doi [\apj] {10.3847/1538-4357/ab017d}, \href
  {https://ui.adsabs.harvard.edu/abs/2019ApJ...872..171L} {872, 171}

\bibitem[\protect\citeauthoryear{{Mart{\'{\i}}n-Pintado}, {de Vicente},
  {Fuente}  \& {Planesas}}{{Mart{\'{\i}}n-Pintado}
  et~al.}{1997}]{Martin-pintado1997}
{Mart{\'{\i}}n-Pintado} J.,  {de Vicente} P.,  {Fuente} A.,   {Planesas} P.,
  1997, \mn@doi [\apjl] {10.1086/310691}, \href
  {http://adsabs.harvard.edu/abs/1997ApJ...482L..45M} {482, L45}

\bibitem[\protect\citeauthoryear{{Mart{\'{\i}}n-Pintado}, {Gaume},
  {Rodr{\'{\i}}guez-Fern{\'a}ndez}, {de Vicente}  \&
  {Wilson}}{{Mart{\'{\i}}n-Pintado} et~al.}{1999}]{Martin-Pintado1999}
{Mart{\'{\i}}n-Pintado} J.,  {Gaume} R.~A.,  {Rodr{\'{\i}}guez-Fern{\'a}ndez}
  N.,  {de Vicente} P.,   {Wilson} T.~L.,  1999, \mn@doi [\apj]
  {10.1086/307399}, \href {http://adsabs.harvard.edu/abs/1999ApJ...519..667M}
  {519, 667}

\bibitem[\protect\citeauthoryear{{Mart{\'\i}n-Pintado}, {Jim{\'e}nez-Serra},
  {Rodr{\'\i}guez-Franco}, {Mart{\'\i}n}  \& {Thum}}{{Mart{\'\i}n-Pintado}
  et~al.}{2005}]{Martin-Pintado2005}
{Mart{\'\i}n-Pintado} J.,  {Jim{\'e}nez-Serra} I.,  {Rodr{\'\i}guez-Franco} A.,
   {Mart{\'\i}n} S.,   {Thum} C.,  2005, \mn@doi [\apjl] {10.1086/432684},
  \href {https://ui.adsabs.harvard.edu/abs/2005ApJ...628L..61M} {628, L61}

\bibitem[\protect\citeauthoryear{{Mart{\'\i}n}, {Mauersberger},
  {Mart{\'\i}n-Pintado}, {Henkel}  \& {Garc{\'\i}a-Burillo}}{{Mart{\'\i}n}
  et~al.}{2006}]{Martin2006}
{Mart{\'\i}n} S.,  {Mauersberger} R.,  {Mart{\'\i}n-Pintado} J.,  {Henkel} C.,
   {Garc{\'\i}a-Burillo} S.,  2006, \mn@doi [\apjs] {10.1086/503297}, \href
  {https://ui.adsabs.harvard.edu/abs/2006ApJS..164..450M} {164, 450}

\bibitem[\protect\citeauthoryear{{Mart{\'{\i}}n}, {Requena-Torres},
  {Mart{\'{\i}}n-Pintado}  \& {Mauersberger}}{{Mart{\'{\i}}n}
  et~al.}{2008}]{Martin2008}
{Mart{\'{\i}}n} S.,  {Requena-Torres} M.~A.,  {Mart{\'{\i}}n-Pintado} J.,
  {Mauersberger} R.,  2008, \mn@doi [\apj] {10.1086/533409}, \href
  {http://adsabs.harvard.edu/abs/2008ApJ...678..245M} {678, 245}

\bibitem[\protect\citeauthoryear{{McMullin}, {Waters}, {Schiebel}, {Young}  \&
  {Golap}}{{McMullin} et~al.}{2007}]{Mcmullin2007}
{McMullin} J.~P.,  {Waters} B.,  {Schiebel} D.,  {Young} W.,   {Golap} K.,
  2007, in {Shaw} R.~A.,  {Hill} F.,   {Bell} D.~J.,  eds,  Astronomical
  Society of the Pacific Conference Series Vol. 376, Astronomical Data Analysis
  Software and Systems XVI. p.~127

\bibitem[\protect\citeauthoryear{{Miao}, {Mehringer}, {Kuan}  \&
  {Snyder}}{{Miao} et~al.}{1995}]{Miao1995}
{Miao} Y.,  {Mehringer} D.~M.,  {Kuan} Y.-J.,   {Snyder} L.~E.,  1995, \mn@doi
  [\apjl] {10.1086/187889}, \href
  {https://ui.adsabs.harvard.edu/abs/1995ApJ...445L..59M} {445, L59}

\bibitem[\protect\citeauthoryear{{Mills}, {Ginsburg}, {Immer}, {Barnes},
  {Wiesenfeld}, {Faure}, {Morris}  \& {Requena-Torres}}{{Mills}
  et~al.}{2018}]{mills2018}
{Mills} E.~A.~C.,  {Ginsburg} A.,  {Immer} K.,  {Barnes} J.~M.,  {Wiesenfeld}
  L.,  {Faure} A.,  {Morris} M.~R.,   {Requena-Torres} M.~A.,  2018, \mn@doi
  [\apj] {10.3847/1538-4357/aae581}, \href
  {http://adsabs.harvard.edu/abs/2018ApJ...868....7M} {868, 7}

\bibitem[\protect\citeauthoryear{{Minh} \& {Irvine}}{{Minh} \&
  {Irvine}}{2006}]{Minh2006}
{Minh} Y.~C.,  {Irvine} W.~M.,  2006, \mn@doi [\na]
  {10.1016/j.newast.2006.03.004}, \href
  {http://adsabs.harvard.edu/abs/2006NewA...11..594M} {11, 594}

\bibitem[\protect\citeauthoryear{{Morris} \& {Serabyn}}{{Morris} \&
  {Serabyn}}{1996}]{Morris1996}
{Morris} M.,  {Serabyn} E.,  1996, \mn@doi [\araa]
  {10.1146/annurev.astro.34.1.645}, \href
  {https://ui.adsabs.harvard.edu/abs/1996ARA\%26A..34..645M} {34, 645}

\bibitem[\protect\citeauthoryear{{M{\"u}ller}, {Thorwirth}, {Roth}  \&
  {Winnewisser}}{{M{\"u}ller} et~al.}{2001}]{Muller2001}
{M{\"u}ller} H.~S.~P.,  {Thorwirth} S.,  {Roth} D.~A.,   {Winnewisser} G.,
  2001, \mn@doi [\aap] {10.1051/0004-6361:20010367}, \href
  {https://ui.adsabs.harvard.edu/abs/2001A\%26A...370L..49M} {370, L49}

\bibitem[\protect\citeauthoryear{{M{\"u}ller}, {Schl{\"o}der}, {Stutzki}  \&
  {Winnewisser}}{{M{\"u}ller} et~al.}{2005}]{Muller2005}
{M{\"u}ller} H.~S.~P.,  {Schl{\"o}der} F.,  {Stutzki} J.,   {Winnewisser} G.,
  2005, \mn@doi [Journal of Molecular Structure]
  {10.1016/j.molstruc.2005.01.027}, \href
  {https://ui.adsabs.harvard.edu/abs/2005JMoSt.742..215M} {742, 215}

\bibitem[\protect\citeauthoryear{{Nagayama}, {Omodaka}, {Handa}, {Toujima},
  {Sofue}, {Sawada}, {Kobayashi}  \& {Koyama}}{{Nagayama}
  et~al.}{2009}]{Nagayama2009}
{Nagayama} T.,  {Omodaka} T.,  {Handa} T.,  {Toujima} H.,  {Sofue} Y.,
  {Sawada} T.,  {Kobayashi} H.,   {Koyama} Y.,  2009, \mn@doi [\pasj]
  {10.1093/pasj/61.5.1023}, \href
  {https://ui.adsabs.harvard.edu/abs/2009PASJ...61.1023N} {61, 1023}

\bibitem[\protect\citeauthoryear{{Ossenkopf} \& {Henning}}{{Ossenkopf} \&
  {Henning}}{1994}]{Ossenkopf1994}
{Ossenkopf} V.,  {Henning} T.,  1994, \aap, \href
  {https://ui.adsabs.harvard.edu/abs/1994A&A...291..943O} {291, 943}

\bibitem[\protect\citeauthoryear{{Ott}, {Wei{\ss}}, {Staveley-Smith}, {Henkel}
  \& {Meier}}{{Ott} et~al.}{2014}]{Ott2014}
{Ott} J.,  {Wei{\ss}} A.,  {Staveley-Smith} L.,  {Henkel} C.,   {Meier} D.~S.,
  2014, \mn@doi [\apj] {10.1088/0004-637X/785/1/55}, \href
  {https://ui.adsabs.harvard.edu/abs/2014ApJ...785...55O} {785, 55}

\bibitem[\protect\citeauthoryear{{Pardo}, {Cernicharo}  \& {Serabyn}}{{Pardo}
  et~al.}{2001}]{Pardo2001}
{Pardo} J.~R.,  {Cernicharo} J.,   {Serabyn} E.,  2001, \mn@doi [IEEE
  Transactions on Antennas and Propagation] {10.1109/8.982447}, \href
  {https://ui.adsabs.harvard.edu/abs/2001ITAP...49.1683P} {49, 1683}

\bibitem[\protect\citeauthoryear{{Pickett}, {Poynter}, {Cohen}, {Delitsky},
  {Pearson}  \& {M{\"u}ller}}{{Pickett} et~al.}{1998}]{Pickett1998}
{Pickett} H.~M.,  {Poynter} R.~L.,  {Cohen} E.~A.,  {Delitsky} M.~L.,
  {Pearson} J.~C.,   {M{\"u}ller} H.~S.~P.,  1998, \mn@doi [J. Quant.
  Spectrosc. & Rad. Transfer] {10.1016/S0022-4073(98)00091-0}, \href
  {http://adsabs.harvard.edu/abs/1998JQSRT..60..883P} {60, 883}

\bibitem[\protect\citeauthoryear{{Pihlstr{\"o}m}, {Sjouwerman}  \&
  {Fish}}{{Pihlstr{\"o}m} et~al.}{2011}]{Pihlstrom2011}
{Pihlstr{\"o}m} Y.~M.,  {Sjouwerman} L.~O.,   {Fish} V.~L.,  2011, \mn@doi
  [\apjl] {10.1088/2041-8205/739/1/L21}, \href
  {https://ui.adsabs.harvard.edu/abs/2011ApJ...739L..21P} {739, L21}

\bibitem[\protect\citeauthoryear{{Pihlstr{\"o}m}, {Sjouwerman}, {Frail},
  {Claussen}, {Mesler}  \& {McEwen}}{{Pihlstr{\"o}m}
  et~al.}{2014}]{Pihlstrom2014}
{Pihlstr{\"o}m} Y.~M.,  {Sjouwerman} L.~O.,  {Frail} D.~A.,  {Claussen} M.~J.,
  {Mesler} R.~A.,   {McEwen} B.~C.,  2014, \mn@doi [\aj]
  {10.1088/0004-6256/147/4/73}, \href
  {https://ui.adsabs.harvard.edu/abs/2014AJ....147...73P} {147, 73}

\bibitem[\protect\citeauthoryear{{Pratap}, {Shute}, {Keane}, {Battersby}  \&
  {Sterling}}{{Pratap} et~al.}{2008}]{Pratap2008}
{Pratap} P.,  {Shute} P.~A.,  {Keane} T.~C.,  {Battersby} C.,   {Sterling} S.,
  2008, \mn@doi [\aj] {10.1088/0004-6256/135/5/1718}, \href
  {https://ui.adsabs.harvard.edu/abs/2008AJ....135.1718P} {135, 1718}

\bibitem[\protect\citeauthoryear{{Rabli} \& {Flower}}{{Rabli} \&
  {Flower}}{2010}]{Rabli2010}
{Rabli} D.,  {Flower} D.~R.,  2010, \mn@doi [\mnras]
  {10.1111/j.1365-2966.2010.16671.x}, \href
  {https://ui.adsabs.harvard.edu/abs/2010MNRAS.406...95R} {406, 95}

\bibitem[\protect\citeauthoryear{{Reid} et~al.,}{{Reid}
  et~al.}{2014}]{Reid2014}
{Reid} M.~J.,  et~al., 2014, \mn@doi [\apj] {10.1088/0004-637X/783/2/130},
  \href {http://adsabs.harvard.edu/abs/2014ApJ...783..130R} {783, 130}

\bibitem[\protect\citeauthoryear{{Requena-Torres}, {Mart{\'{\i}}n-Pintado},
  {Rodr{\'{\i}}guez-Franco}, {Mart{\'{\i}}n}, {Rodr{\'{\i}}guez-Fern{\'a}ndez}
  \& {de Vicente}}{{Requena-Torres} et~al.}{2006}]{Requena-torres2006}
{Requena-Torres} M.~A.,  {Mart{\'{\i}}n-Pintado} J.,  {Rodr{\'{\i}}guez-Franco}
  A.,  {Mart{\'{\i}}n} S.,  {Rodr{\'{\i}}guez-Fern{\'a}ndez} N.~J.,   {de
  Vicente} P.,  2006, \mn@doi [\aap] {10.1051/0004-6361:20065190}, \href
  {http://adsabs.harvard.edu/abs/2006A\%26A...455..971R} {455, 971}

\bibitem[\protect\citeauthoryear{{Requena-Torres}, {Mart{\'{\i}}n-Pintado},
  {Mart{\'{\i}}n}  \& {Morris}}{{Requena-Torres}
  et~al.}{2008}]{Requena-torres2008}
{Requena-Torres} M.~A.,  {Mart{\'{\i}}n-Pintado} J.,  {Mart{\'{\i}}n} S.,
  {Morris} M.~R.,  2008, \mn@doi [\apj] {10.1086/523627}, \href
  {http://adsabs.harvard.edu/abs/2008ApJ...672..352R} {672, 352}

\bibitem[\protect\citeauthoryear{{Rivilla} et~al.,}{{Rivilla}
  et~al.}{2018}]{Rivilla2018}
{Rivilla} V.~M.,  et~al., 2018, \mn@doi [\mnras] {10.1093/mnrasl/slx208}, \href
  {http://adsabs.harvard.edu/abs/2018MNRAS.475L..30R} {475, L30}

\bibitem[\protect\citeauthoryear{{Rivilla} et~al.,}{{Rivilla}
  et~al.}{2019}]{Rivilla2019}
{Rivilla} V.~M.,  et~al., 2019, \mn@doi [\mnras] {10.1093/mnrasl/sly228}, \href
  {http://adsabs.harvard.edu/abs/2019MNRAS.483L.114R} {483, L114}

\bibitem[\protect\citeauthoryear{{Rodr{\'{\i}}guez-Fern{\'a}ndez},
  {Mart{\'{\i}}n-Pintado}, {de Vicente}, {Fuente}, {H{\"u}ttemeister}, {Wilson}
   \& {Kunze}}{{Rodr{\'{\i}}guez-Fern{\'a}ndez}
  et~al.}{2000}]{Rodriguez-fernandez2000}
{Rodr{\'{\i}}guez-Fern{\'a}ndez} N.~J.,  {Mart{\'{\i}}n-Pintado} J.,  {de
  Vicente} P.,  {Fuente} A.,  {H{\"u}ttemeister} S.,  {Wilson} T.~L.,   {Kunze}
  D.,  2000, \aap, \href {http://adsabs.harvard.edu/abs/2000A\%26A...356..695R}
  {356, 695}

\bibitem[\protect\citeauthoryear{{Rodr{\'{\i}}guez-Fern{\'a}ndez},
  {Mart{\'{\i}}n-Pintado}, {Fuente}  \&
  {Wilson}}{{Rodr{\'{\i}}guez-Fern{\'a}ndez}
  et~al.}{2004}]{Rodriguez-fernandez2004}
{Rodr{\'{\i}}guez-Fern{\'a}ndez} N.~J.,  {Mart{\'{\i}}n-Pintado} J.,  {Fuente}
  A.,   {Wilson} T.~L.,  2004, \mn@doi [\aap] {10.1051/0004-6361:20041370},
  \href {http://adsabs.harvard.edu/abs/2004A\%26A...427..217R} {427, 217}

\bibitem[\protect\citeauthoryear{{Salii}, {Sobolev}  \& {Kalinina}}{{Salii}
  et~al.}{2002}]{Salii2002}
{Salii} S.~V.,  {Sobolev} A.~M.,   {Kalinina} N.~D.,  2002, \mn@doi [Astronomy
  Reports] {10.1134/1.1529254}, \href
  {https://ui.adsabs.harvard.edu/abs/2002ARep...46..955S} {46, 955}

\bibitem[\protect\citeauthoryear{{Sato}, {Hasegawa}, {Whiteoak}  \&
  {Miyawaki}}{{Sato} et~al.}{2000}]{Sato2000}
{Sato} F.,  {Hasegawa} T.,  {Whiteoak} J.~B.,   {Miyawaki} R.,  2000, \mn@doi
  [\apj] {10.1086/308856}, \href
  {http://adsabs.harvard.edu/abs/2000ApJ...535..857S} {535, 857}

\bibitem[\protect\citeauthoryear{{Sault}, {Teuben}  \& {Wright}}{{Sault}
  et~al.}{1995}]{sault1995}
{Sault} R.~J.,  {Teuben} P.~J.,   {Wright} M.~C.~H.,  1995, in {Shaw} R.~A.,
  {Payne} H.~E.,   {Hayes} J.~J.~E.,  eds,  Astronomical Society of the Pacific
  Conference Series Vol. 77, Astronomical Data Analysis Software and Systems
  IV. p.~433 (\mn@eprint {} {astro-ph/0612759})

\bibitem[\protect\citeauthoryear{{Schmiedeke} et~al.,}{{Schmiedeke}
  et~al.}{2016}]{Schmiedeke2016}
{Schmiedeke} A.,  et~al., 2016, \mn@doi [\aap] {10.1051/0004-6361/201527311},
  \href {https://ui.adsabs.harvard.edu/abs/2016A&A...588A.143S} {588, A143}

\bibitem[\protect\citeauthoryear{{Shima}, {Tasker}  \& {Habe}}{{Shima}
  et~al.}{2016}]{Shima2016}
{Shima} K.,  {Tasker} E.~J.,   {Habe} A.,  2016, in {Jablonka} P.,  {Andr{\'e}}
  P.,   {van der Tak} F.,  eds,  IAU Symposium Vol. 315, From Interstellar
  Clouds to Star-Forming Galaxies: Universal Processes?. p.~E72,
  \mn@doi{10.1017/S1743921316008346}

\bibitem[\protect\citeauthoryear{{Sjouwerman}, {Pihlstr{\"o}m}  \&
  {Fish}}{{Sjouwerman} et~al.}{2010}]{Sjouwerman2010}
{Sjouwerman} L.~O.,  {Pihlstr{\"o}m} Y.~M.,   {Fish} V.~L.,  2010, \mn@doi
  [\apjl] {10.1088/2041-8205/710/2/L111}, \href
  {https://ui.adsabs.harvard.edu/abs/2010ApJ...710L.111S} {710, L111}

\bibitem[\protect\citeauthoryear{{Slane}, {Bykov}, {Ellison}, {Dubner}  \&
  {Castro}}{{Slane} et~al.}{2015}]{Slane2015}
{Slane} P.,  {Bykov} A.,  {Ellison} D.~C.,  {Dubner} G.,   {Castro} D.,  2015,
  \mn@doi [\ssr] {10.1007/s11214-014-0062-6}, \href
  {https://ui.adsabs.harvard.edu/abs/2015SSRv..188..187S} {188, 187}

\bibitem[\protect\citeauthoryear{{Sobolev}}{{Sobolev}}{1992}]{Sobolev1992}
{Sobolev} A.~M.,  1992, \sovast, \href
  {https://ui.adsabs.harvard.edu/abs/1992SvA....36..590S} {36, 590}

\bibitem[\protect\citeauthoryear{{Takahira}, {Tasker}  \& {Habe}}{{Takahira}
  et~al.}{2014}]{Takahira2014}
{Takahira} K.,  {Tasker} E.~J.,   {Habe} A.,  2014, \mn@doi [\apj]
  {10.1088/0004-637X/792/1/63}, \href
  {https://ui.adsabs.harvard.edu/abs/2014ApJ...792...63T} {792, 63}

\bibitem[\protect\citeauthoryear{{Torii} et~al.,}{{Torii}
  et~al.}{2017}]{Torii2017}
{Torii} K.,  et~al., 2017, \mn@doi [\apj] {10.3847/1538-4357/835/2/142}, \href
  {https://ui.adsabs.harvard.edu/abs/2017ApJ...835..142T} {835, 142}

\bibitem[\protect\citeauthoryear{{Tsuboi}, {Miyazaki}  \& {Uehara}}{{Tsuboi}
  et~al.}{2015}]{Tsuboi2015}
{Tsuboi} M.,  {Miyazaki} A.,   {Uehara} K.,  2015, \mn@doi [\pasj]
  {10.1093/pasj/psv058}, \href
  {http://adsabs.harvard.edu/abs/2015PASJ...67...90T} {67, 90}

\bibitem[\protect\citeauthoryear{{Van der Tak}, {Black}, {Sch{\"o}ier},
  {Jansen}  \& {van Dishoeck}}{{Van der Tak} et~al.}{2007}]{Vandetak2007}
{Van der Tak} F.~F.~S.,  {Black} J.~H.,  {Sch{\"o}ier} F.~L.,  {Jansen} D.~J.,
   {van Dishoeck} E.~F.,  2007, \mn@doi [\aap] {10.1051/0004-6361:20066820},
  \href {https://ui.adsabs.harvard.edu/abs/2007A&A...468..627V} {468, 627}

\bibitem[\protect\citeauthoryear{{Voronkov}, {Brooks}, {Sobolev}, {Ellingsen},
  {Ostrovskii}  \& {Caswell}}{{Voronkov} et~al.}{2006}]{Voronkov2006}
{Voronkov} M.~A.,  {Brooks} K.~J.,  {Sobolev} A.~M.,  {Ellingsen} S.~P.,
  {Ostrovskii} A.~B.,   {Caswell} J.~L.,  2006, \mn@doi [\mnras]
  {10.1111/j.1365-2966.2006.11047.x}, \href
  {https://ui.adsabs.harvard.edu/abs/2006MNRAS.373..411V} {373, 411}

\bibitem[\protect\citeauthoryear{{Voronkov}, {Caswell}, {Britton}, {Green},
  {Sobolev}  \& {Ellingsen}}{{Voronkov} et~al.}{2010}]{Voronkov2010a}
{Voronkov} M.~A.,  {Caswell} J.~L.,  {Britton} T.~R.,  {Green} J.~A.,
  {Sobolev} A.~M.,   {Ellingsen} S.~P.,  2010, \mn@doi [\mnras]
  {10.1111/j.1365-2966.2010.17222.x}, \href
  {https://ui.adsabs.harvard.edu/abs/2010MNRAS.408..133V} {408, 133}

\bibitem[\protect\citeauthoryear{{Voronkov}, {Caswell}, {Ellingsen}, {Breen},
  {Britton}, {Green}, {Sobolev}  \& {Walsh}}{{Voronkov}
  et~al.}{2012}]{Voronkov2012}
{Voronkov} M.~A.,  {Caswell} J.~L.,  {Ellingsen} S.~P.,  {Breen} S.~L.,
  {Britton} T.~R.,  {Green} J.~A.,  {Sobolev} A.~M.,   {Walsh} A.~J.,  2012, in
  {Booth} R.~S.,  {Vlemmings} W. H.~T.,   {Humphreys} E. M.~L.,  eds,  IAU
  Symposium Vol. 287, Cosmic Masers - from OH to H0. pp 433--440 (\mn@eprint
  {arXiv} {1203.5492}), \mn@doi{10.1017/S174392131200748X}

\bibitem[\protect\citeauthoryear{{Voronkov}, {Caswell}, {Ellingsen}, {Green}
  \& {Breen}}{{Voronkov} et~al.}{2014}]{Voronkov2014}
{Voronkov} M.~A.,  {Caswell} J.~L.,  {Ellingsen} S.~P.,  {Green} J.~A.,
  {Breen} S.~L.,  2014, \mn@doi [\mnras] {10.1093/mnras/stu116}, \href
  {https://ui.adsabs.harvard.edu/abs/2014MNRAS.439.2584V} {439, 2584}

\bibitem[\protect\citeauthoryear{{Yusef-Zadeh}, {Wardle}, {Rho}  \&
  {Sakano}}{{Yusef-Zadeh} et~al.}{2003}]{Yusef-Zadeh2003}
{Yusef-Zadeh} F.,  {Wardle} M.,  {Rho} J.,   {Sakano} M.,  2003, \mn@doi [\apj]
  {10.1086/345932}, \href
  {https://ui.adsabs.harvard.edu/abs/2003ApJ...585..319Y} {585, 319}

\bibitem[\protect\citeauthoryear{{Yusef-Zadeh} et~al.,}{{Yusef-Zadeh}
  et~al.}{2013}]{Yusef-Zadeh2013}
{Yusef-Zadeh} F.,  et~al., 2013, \mn@doi [\apj] {10.1088/0004-637X/762/1/33},
  \href {https://ui.adsabs.harvard.edu/abs/2013ApJ...762...33Y} {762, 33}

\bibitem[\protect\citeauthoryear{{Zeng} et~al.,}{{Zeng}
  et~al.}{2018}]{Zeng2018}
{Zeng} S.,  et~al., 2018, \mn@doi [\mnras] {10.1093/mnras/sty1174}, \href
  {http://adsabs.harvard.edu/abs/2018MNRAS.478.2962Z} {478, 2962}

\bibitem[\protect\citeauthoryear{{de Vicente} et~al.,}{{de Vicente}
  et~al.}{2016}]{deVicente2016}
{de Vicente} P.,  et~al., 2016, \mn@doi [\aap] {10.1051/0004-6361/201527174},
  \href {https://ui.adsabs.harvard.edu/abs/2016A&A...589A..74D} {589, A74}

\makeatother
\end{thebibliography}

%\begin{thebibliography}{99}
%\end{thebibliography}

%%%%%%%%%%%%%%%%%%%%%%%%%%%%%%%%%%%%%%%%%%%%%%%%%%

%%%%%%%%%%%%%%%%% APPENDICES %%%%%%%%%%%%%%%%%%%%%

\appendix
\label{appendix}
\section{Integrated molecular emission map and position-velocity map of HNCO, SiO, and SO}

%If you want to present additional material which would interrupt the flow of the main paper, it can be placed in an Appendix which appears after the list of references.
%%%%%%%%%%%%%%%%%%%%%%%%%%%%%%%%%%%%%%%%%%
%%%%%%%%%%%%%%%%%%%%%%%%%%%%%%%%%%%%%%%%%%
\clearpage

\begin{figure*}
    \centering
    \includegraphics[width=\textwidth]{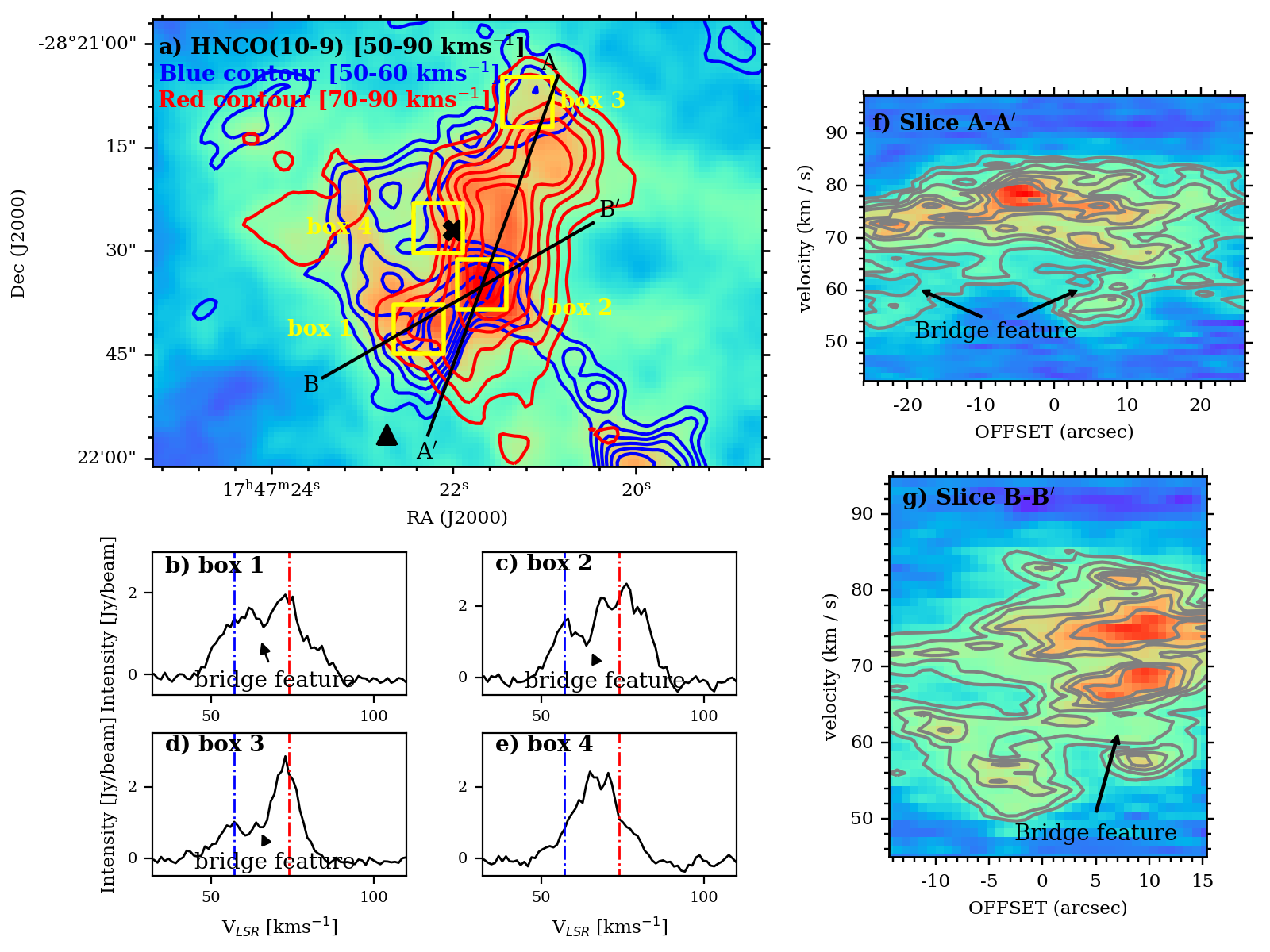}
    %\vspace{-1cm}
    \caption{HNCO(10-9). Caption is the same as Figure \ref{fig:molecular_pv}.}
    \label{fig:molecular_pv_HNCO}
\end{figure*}

\begin{figure*}
    \centering
    \includegraphics[width=\textwidth]{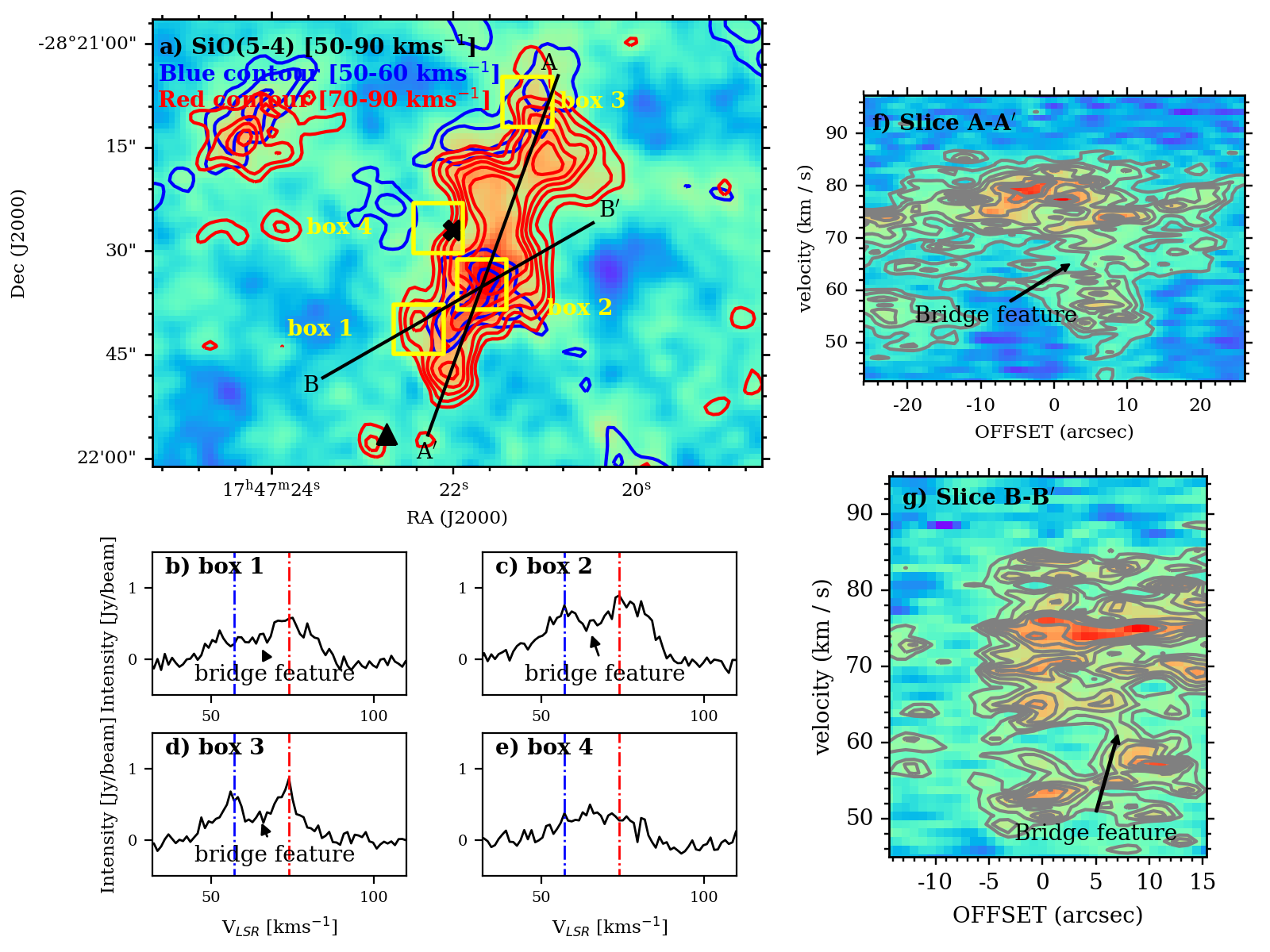}
    %\vspace{-1cm}
    \caption{SiO(5-4). Caption is the same as Figure \ref{fig:molecular_pv}.}
    \label{fig:molecular_pv_SiO}
\end{figure*}

\begin{figure*}
    \centering
    \includegraphics[width=\textwidth]{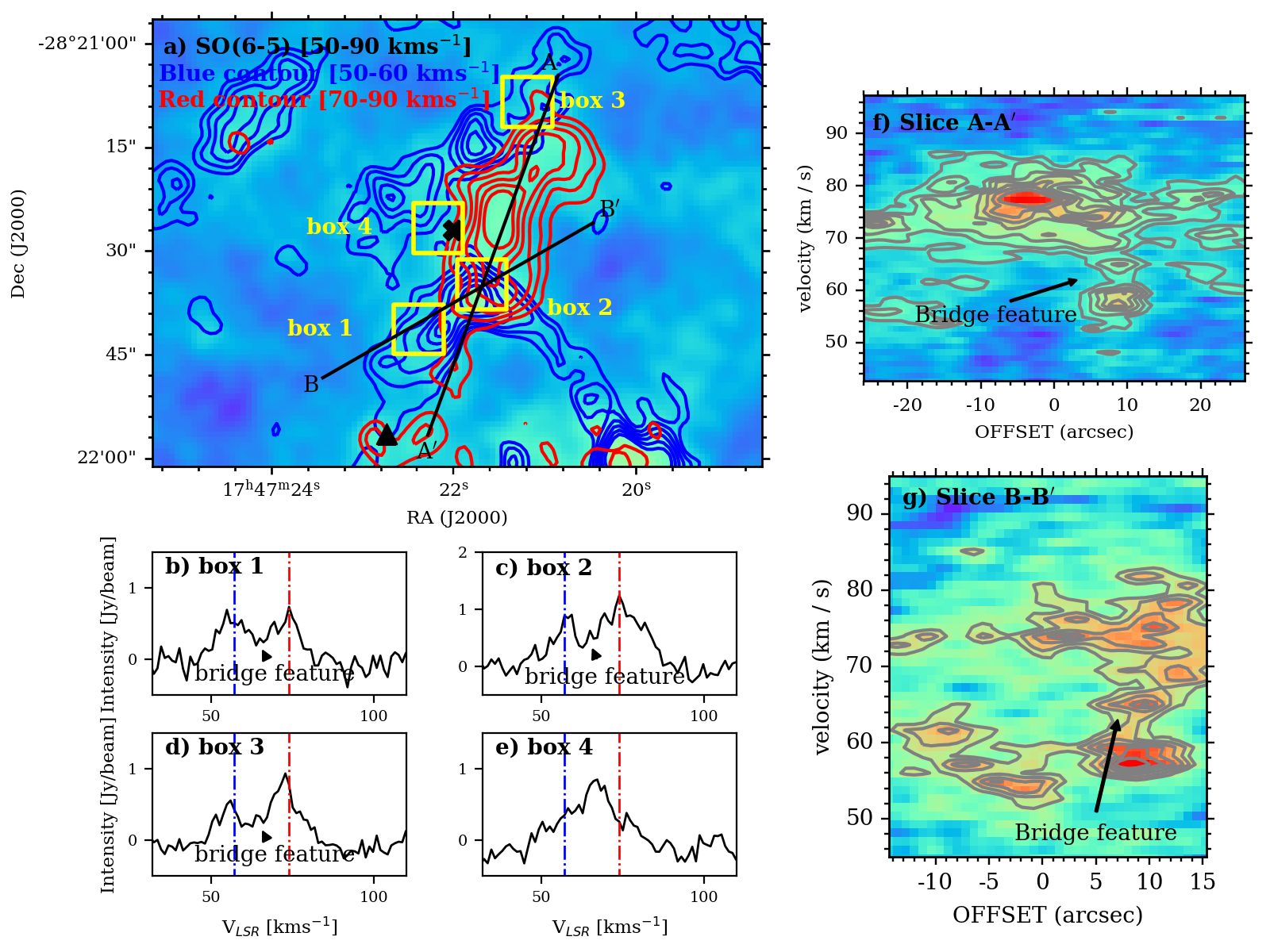}
    %\vspace{-1cm}
    \caption{SO(6-5). Caption is the same as Figure \ref{fig:molecular_pv}.}
    \label{fig:molecular_pv_SO}
\end{figure*}
%%%%%%%%%%%%%%%%%%%%%%%%%%%%%%%%%%%%%%%%%%
%%%%%%%%%%%%%%%%%%%%%%%%%%%%%%%%%%%%%%%%%%

%%%%%%%%%%%%%%%%%%%%%%%%%%%%%%%%%%%%%%%%%%%%%%%%%%

% Don't change these lines
\bsp	% typesetting comment
\label{lastpage}
\end{document}